\documentclass[12pt]{article}
\usepackage{amsmath,float,amssymb,amsthm,amsxtra,overpic,bbm,bm,epsfig}
\def\thefootnote{\fnsymbol{footnote}}

\begin{document}

\vspace{0.2cm}
\begin{center}
{\large\bf Neutrino mass ordering and
$\mu$-$\tau$ reflection symmetry breaking}
\end{center}
	
\vspace{0.1cm}
	
\begin{center}
{\bf Zhi-zhong Xing}$^{1, 2, 3}$ \footnote{E-mail: xingzz@ihep.ac.cn}
~and~ {\bf Jing-yu Zhu}$^{1,2}$ \footnote{E-mail: zhujingyu@ihep.ac.cn} \\
{\small $^{1}$Institute of High Energy Physics, Chinese Academy of
Sciences, Beijing 100049, China \\
$^2$School of Physical Sciences, University of Chinese Academy of
Sciences, Beijing 100049, China \\
$^3$Center for High Energy Physics, Peking University, Beijing
100080, China }
\end{center}
	
\vspace{1.5cm}
	
\begin{abstract}
If the neutrino mass spectrum turns out to be 
$m^{}_3 < m^{}_1 < m^{}_2$, one may choose to relabel it as 
$m^{\prime}_1 < m^{\prime}_2 < m^{\prime}_3$ such that all the masses
of fundamental fermions with the same electrical charges are in order.
In this case the columns of the $3\times 3$ lepton flavor mixing
matrix $U$ should be reordered accordingly,
and the resulting pattern $U^\prime$ may involve one or two
large mixing angles in the standard parametrization
or its variations. Since the Majorana neutrino mass
matrix keeps unchanged in such a mass relabeling, 
a possible $\mu$-$\tau$ reflection symmetry is
respected in this connection and its breaking effects are
model-independently constrained at the $3\sigma$ level by using
current experimental data.
\end{abstract}
	
\begin{flushleft}
\hspace{0.8cm} PACS number(s): 14.60.Pq, 13.10.+q, 25.30.Pt \\
\end{flushleft}
\def\thefootnote{\arabic{footnote}}
\setcounter{footnote}{0}
\newpage
\section{Introduction}
In a simple extension of the standard electroweak model
which can generate finite but tiny masses for the three originally
massless neutrinos, the phenomenon of lepton flavor mixing measures
a nontrivial mismatch between the mass and flavor eigenstates
of the charged leptons and neutrinos. Similar to the dynamics
of quark flavor mixing, the conjecture that the lepton fields interact
simultaneously with the scalar and gauge fields leads to both
lepton flavor mixing and CP violation in the standard three-flavor scheme \cite{XZ}.
The $3\times 3$ lepton and quark mixing matrices which manifest themselves
in the weak charged-current interactions are referred to as the
Pontecorvo-Maki-Nakagawa-Sakata (PMNS) matrix $U$ \cite{PMNS} and
the Cabibbo-Kobayashi-Maskawa (CKM) matrix $V$ \cite{CKM}, respectively:
\begin{eqnarray}
-{\cal L}^{}_{\rm cc} & = & \frac{g}{\sqrt{2}} \left[
\overline{\left(e ~~ \mu ~~ \tau \right)^{}_{\rm L}} \ \gamma^\mu \
U \left(
    \begin{matrix}
        \nu^{}_1 \cr \nu^{}_2 \cr \nu^{}_3\cr
    \end{matrix}
\right)_{\rm L} W^-_\mu ~ + ~ \overline{\left(u ~~ c ~~ t
\right)^{}_{\rm L}} \ \gamma^\mu \ V \left(
    \begin{matrix}
        d \cr s \cr b\cr
    \end{matrix}
\right)_{\rm L} W^+_\mu \right] + {\rm h.c.}
\end{eqnarray}
with all the fermion fields being the mass eigenstates. By
convention $U$ and $V$ are defined to be associated with $W^-$ and
$W^+$, respectively. In Eq. (1) the charged leptons and quarks
with the same electric charges all have the ``normal"
\footnote{Here ``normal" means that the charged fermions in
the first family are the lightest and those in the third family
are the heaviest when their masses are renormalized to a
common energy scale, as shown in Figure 1.}
and strong mass hierarchies \cite{PDG},
\begin{eqnarray}
m^{}_e \ll m^{}_\mu \ll m^{}_\tau \; , ~~~
m^{}_u \ll m^{}_c \ll m^{}_t \; , ~~~
m^{}_d \ll m^{}_s \ll m^{}_b \; .
\end{eqnarray}
But for the time being it remains unclear whether the three neutrinos
also have a normal mass
ordering $m^{}_1 < m^{}_2 < m^{}_3$ or not. Now that $m^{}_1 < m^{}_2$
has been fixed from the solar neutrino oscillation data by the 
$0 \leq \theta^{}_{12} \leq \pi/4$ convention and with
the help of matter effects \cite{Hagiwara},
the only possible abnormal neutrino mass ordering
is $m^{}_3 < m^{}_1 < m^{}_2$, which has been referred to as the
``inverted" mass ordering. Needless to say, the neutrino mass ordering
is one of the central concerns in today's neutrino physics.
A combination of current atmospheric (Super-Kamiokande) and
accelerator-based (T2K and NO$\nu$A) neutrino oscillation data
preliminarily favors the normal neutrino mass ordering up to
the $2\sigma$ level \cite{Lisi,T2K}. But one has to be very cautious and
open-minded at this stage, because only the next-generation neutrino
oscillation experiments can really pin down the true neutrino mass ordering
\cite{Schwetz}.

If the masses of three neutrinos end up in an inverted ordering,
which is quite different as compared with the mass spectra of
their nine charged partners in the standard model (see Figure 1 for illustration),
one will have to explain or understand what is behind this ``anomaly" from
a theoretical point of view. From a purely phenomenological point of
view, we may simply choose to relabel the three neutrino mass eigenstates
in Eq. (1),
\begin{eqnarray}
\left(
    \begin{matrix}
        \nu^\prime_1 \cr \nu^\prime_2 \cr \nu^\prime_3\cr
    \end{matrix}
\right) = S \left(
    \begin{matrix}
        \nu^{}_1 \cr \nu^{}_2 \cr \nu^{}_3\cr
    \end{matrix}
\right) ~~ {\rm with} ~~~ S = \left(
    \begin{matrix}
    0 & 0 & 1 \cr 1 & 0 & 0 \cr 0 & 1 & 0 \cr
    \end{matrix}
\right) \; ,
\end{eqnarray}
and thus make the mass spectrum become ``normal":
$m^\prime_1 \equiv m^{}_3 < m^\prime_2 \equiv m^{}_1 < m^\prime_3
\equiv m^{}_2$, as shown in Figure 2. The invariance of the weak
charged-current interactions (i.e., ${\cal L}^{}_{\rm cc}$) under such a
transformation requires that the PMNS matrix $U$ transform accordingly,
\begin{eqnarray}
U^\prime = \left(
\begin{matrix}
    U^\prime_{e1} & U^\prime_{e2} & U^\prime_{e3} \cr
    U^\prime_{\mu 1} & U^\prime_{\mu 2} & U^\prime_{\mu 3} \cr
    U^\prime_{\tau 1} & U^\prime_{\tau 2} & U^\prime_{\tau 3} \cr
\end{matrix}
\right) = U S^{-1} = \left(
\begin{matrix}
    U^{}_{e3} & U^{}_{e1} & U^{}_{e2} \cr
    U^{}_{\mu 3} & U^{}_{\mu 1} & U^{}_{\mu 2} \cr
    U^{}_{\tau 3} & U^{}_{\tau 1} & U^{}_{\tau 2} \cr
\end{matrix}
\right) \; ,
\end{eqnarray}
implying a striking change of the pattern of lepton flavor mixing.
Provided $U^\prime$ is parametrized in the same
way as $U$, the corresponding parameters must take different
values. We shall show that $U^\prime$ may involve one or two
large flavor mixing angles in the standard parametrization
or its variations. We shall also illustrate that such a 
mass relabeling does not change the texture of the
Majorana neutrino mass matrix $M^{}_\nu$ by taking into account
a possible $\mu$-$\tau$ reflection symmetry, and model-independently
constrain its breaking effects at the $3\sigma$ level with the help
of current neutrino oscillation data.
\begin{figure*}[t]
\centering\includegraphics[width=16cm]{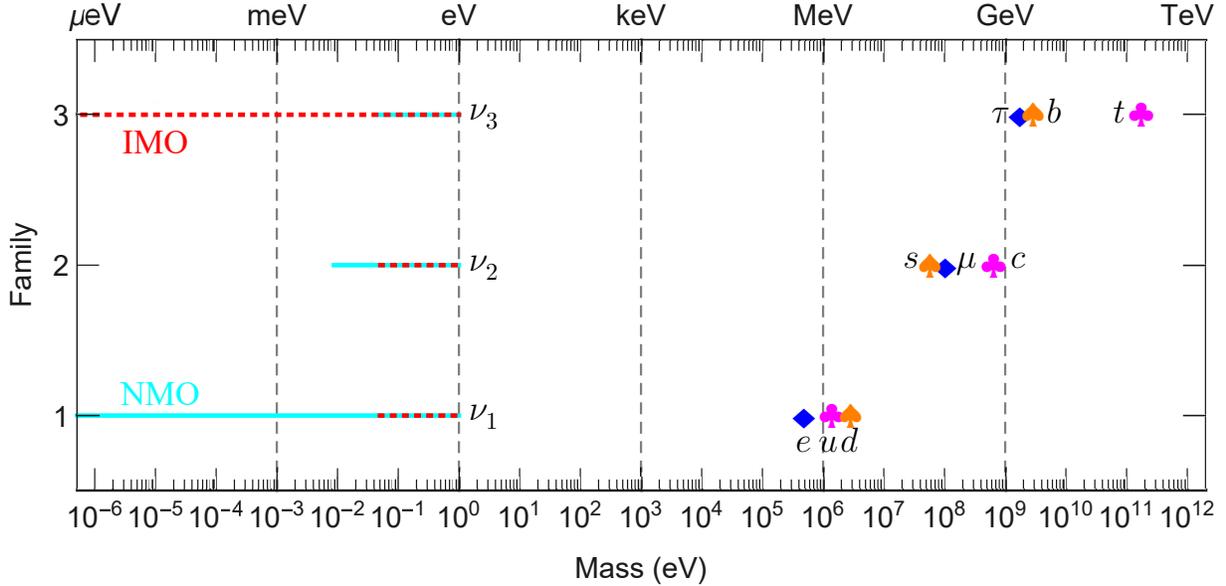}
\caption{A schematic illustration of the
fermion mass spectra at the electroweak scale $M^{}_Z$, where
the allowed ranges of three neutrino masses are quoted
from Ref. \cite{Lisi}, and the typical values of three charged-lepton
masses and six quark masses are quoted from Ref. \cite{XZZ}. As
for $\nu^{}_1$, $\nu^{}_2$ and $\nu^{}_3$, the horizontal cyan lines and red
dashed lines stand for the normal mass ordering (NMO) and inverted
mass ordering (IMO) cases, respectively. Note that the lower limit of
$m^{}_1$ in the NMO case or that of $m^{}_3$ in the IMO case can
be extended to zero.}
\end{figure*}
\begin{figure*}[t]
\centering\includegraphics[width=16cm]{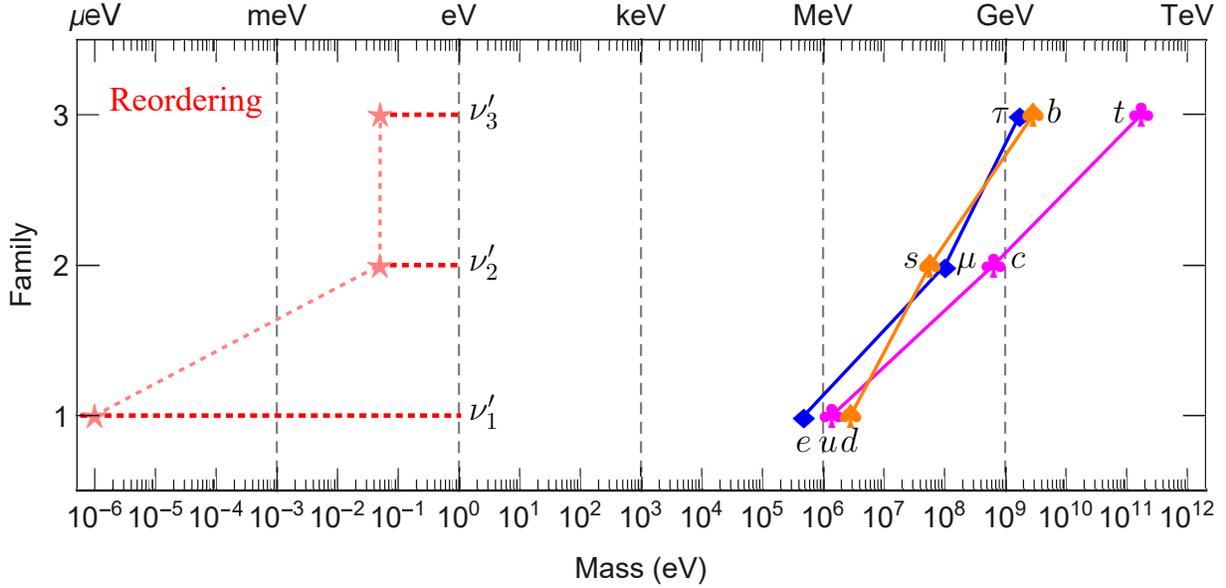}
\caption{A reordering of the IMO case in Figure 1 is likely to
make the neutrino mass spectrum look more or less similar to the mass
spectra of the charged fermions. Here we have shown a simple example
for illustration, by taking $m^{\prime}_1 \sim 10^{-6}$ eV and
$m^{\prime}_2 \simeq m^{\prime}_3 \sim 0.05$ eV.}
\end{figure*}

Although relabeling $m^{}_3 < m^{}_1 < m^{}_2$ as 
$m^\prime_1 < m^\prime_2 < m^\prime_3$ does not
add any new physical content, we stress that it represents a phenomenologically
alternative and interesting way to describe the neutrino mass spectrum and the
lepton flavor mixing pattern, especially when they are compared with (and
even unified with) the ``normal" quark mass spectra and the CKM quark
flavor mixing matrix. That is why we intend to
look into such an option in some detail, and hope that this kind of
study may help make the underlying physics more transparent in some
explicit model-building exercises towards deeper understanding of the
origins of fermion masses and flavor mixing.

\section{Possible patterns of $U$ and $U^\prime$}

Let us focus on the intriguing case that the three massive
neutrinos are of the Majorana nature \cite{M}. In this case the effective
neutrino mass term reads
\begin{eqnarray}
-{\cal L}_{\rm mass} = \frac{1}{2} \overline{ \left(
        \nu^{}_e ~~ \nu^{}_\mu  ~~\nu^{}_\tau
\right)^{}_{\rm L}}
\ M^{}_\nu \left(
    \begin{matrix}
        \nu^{\rm c}_e \cr \nu^{\rm c}_\mu \cr \nu^{\rm c}_\tau\cr
    \end{matrix}
\right)_{\rm R} + ~{\rm h.c.} \; ,
\end{eqnarray}
where the superscript ``$\rm c$" denotes the charge conjugation, and
the $3\times 3$ Majorana mass matrix $M^{}_\nu$ is symmetric and can be expressed as
\begin{eqnarray}
M^{}_\nu = \left(
\begin{matrix}
\langle m\rangle^{}_{ee} &
\langle m\rangle^{}_{e\mu} & \langle m\rangle^{}_{e\tau} \cr
\langle m\rangle^{}_{e\mu} & \langle m\rangle^{}_{\mu\mu} &
\langle m\rangle^{}_{\mu\tau} \cr \langle m\rangle^{}_{e\tau} &
\langle m\rangle^{}_{\mu\tau} & \langle m\rangle^{}_{\tau\tau} \cr
\end{matrix}
\right) \; .
\end{eqnarray}
By diagonalizing $M^{}_\nu$ one may transform the neutrino flavor eigenstates
$(\nu^{}_e, \nu^{}_\mu, \nu^{}_\tau)$ into the neutrino mass eigenstates
$(\nu^{}_1, \nu^{}_2, \nu^{}_3)$ or $(\nu^{\prime}_1, \nu^{\prime}_2, \nu^{\prime}_3)$.
In the basis where the flavor eigenstates of three charged leptons
are identified with their mass eigenstates, the texture of
$M^{}_\nu$ can be reconstructed in terms of three neutrino
masses and six flavor mixing parameters as follows:
\begin{eqnarray}
M^{}_\nu = U \widehat{M}^{}_\nu U^{\rm T}
= U^\prime S \widehat{M}^{}_\nu S^{\rm T} U^{\prime\rm T}
= U^\prime \widehat{M}^{\prime}_\nu U^{\prime\rm T} \; ,
\end{eqnarray}
where ``$\rm T$" denotes the transpose,
$\widehat{M}^{}_\nu \equiv {\rm Diag}\{m^{}_1, m^{}_2, m^{}_3\}$
and $\widehat{M}^{\prime}_\nu \equiv
{\rm Diag}\{m^{\prime}_1, m^{\prime}_2, m^{\prime}_3\} =
{\rm Diag}\{m^{}_3, m^{}_1, m^{}_2\}$.
Eq. (7) tells us that $M^{}_\nu$ keeps invariant
under the neutrino mass relabeling. This observation is of course
expectable because any physics must be unchanged by reordering the
neutrino mass spectrum itself. It implies that a possible flavor symmetry
of $M^{}_\nu$ will give rise to the same constraints on the patterns
of $U$ and $U^\prime$.

To see the above point of view in a more transparent way and illustrate
the relevant physics, let us impose the flavor transformation
\begin{eqnarray}
\nu^{}_{e \rm L} \longleftrightarrow \nu^{\rm c}_{e \rm R} \; , ~~~
\nu^{}_{\mu \rm L} \longleftrightarrow \nu^{\rm c}_{\tau \rm R} \; , ~~~
\nu^{}_{\tau \rm L} \longleftrightarrow \nu^{\rm c}_{\mu \rm R} \;
\end{eqnarray}
on the neutrino mass term ${\cal L}^{}_{\rm mass}$ in Eq. (5) and
require the latter to be invariant. It is easy to see that the invariance of
${\cal L}^{}_{\rm mass}$ under such a $\mu$-$\tau$ reflection
transformation \cite{HS} cannot hold unless the elements of $M^{}_\nu$
satisfy the four conditions \cite{XZreview}
\begin{eqnarray}
\langle m\rangle^{}_{ee} = \langle m\rangle^*_{ee} \; , ~~~
\langle m\rangle^{}_{e \mu} = \langle m\rangle^*_{e \tau} \; , ~~~
\langle m\rangle^{}_{\mu\mu} = \langle m\rangle^*_{\tau\tau} \; , ~~~
\langle m\rangle^{}_{\mu\tau} = \langle m\rangle^*_{\mu\tau} \; .
\end{eqnarray}
Such a special texture of $M^{}_\nu$ can be diagonalized by
the PMNS matrix
\footnote{In this parametrization
$c^{}_{ij} \equiv \cos\theta^{}_{ij}$ and $s^{}_{ij} \equiv
\sin\theta^{}_{ij}$ (for $ij = 12, 13, 23$) are defined, and
all the three mixing angles are arranged to lie in the first
quadrant. As for the CP-violating phases, $\delta$ varies
from $0$ to $2\pi$, but $\phi$ and $\varphi$ are only allowed
to vary between $0$ and $\pi$. Note that the sign convention of
$U^{}_{\mu i}$ (for $i=1,2,3$) in Eq. (10) is different from
that advocated by the Particle Data Group \cite{PDG},
because the latter will lead to
$\langle m\rangle^{}_{e \mu} = -\langle m\rangle^*_{e \tau}$
which is in conflict with
$\langle m\rangle^{}_{e \mu} = \langle m\rangle^*_{e \tau}$
given in Eq. (9). In Eq. (12) the flavor mixing matrix $U^\prime$ is required
to take the same new sign convention as $U$.}
\begin{eqnarray}
U = \left(
\begin{matrix}
c^{}_{12} c^{}_{13} & s^{}_{12} c^{}_{13} &
s^{}_{13} e^{-{\rm i} \delta} \cr s^{}_{12} c^{}_{23} + c^{}_{12}
s^{}_{13} s^{}_{23} e^{{\rm i} \delta} & -c^{}_{12} c^{}_{23} +
s^{}_{12} s^{}_{13} s^{}_{23} e^{{\rm i} \delta} & -c^{}_{13}
s^{}_{23} \cr s^{}_{12} s^{}_{23} - c^{}_{12} s^{}_{13} c^{}_{23}
e^{{\rm i} \delta} & -c^{}_{12} s^{}_{23} - s^{}_{12} s^{}_{13}
c^{}_{23} e^{{\rm i} \delta} & c^{}_{13} c^{}_{23} \cr
\end{matrix} \right) \hspace{-0.17cm}
\left( \begin{matrix}
e^{{\rm i} \phi} & 0 & 0 \cr 0 & e^{{\rm i} \varphi} & 0 \cr 0 & 0 &
1 \cr \end{matrix} \right)
\end{eqnarray}
via the transformation $U^\dagger M^{}_\nu U^* = \widehat{M}^{}_\nu$,
where
\begin{eqnarray}
\theta^{}_{23} = \frac{\pi}{4} \; , ~~~ \delta = \frac{\pi}{2} ~ {\rm or} ~
\frac{3\pi}{2} \; , ~~~ \phi = 0 ~{\rm or} ~ \frac{\pi}{2} \; , ~~~
\varphi = 0 ~ {\rm or} ~ \frac{\pi}{2} \; .
\end{eqnarray}
In view of Eq. (7), it is easy to show that the texture of $M^{}_\nu$
constrained by Eq. (9) can also be diagonalized by
\begin{eqnarray}
U^\prime = \left(
\begin{matrix}
c^{\prime}_{12} c^{\prime}_{13} &
s^{\prime}_{12} c^{\prime}_{13} &
s^{\prime}_{13} e^{-{\rm i} \delta^\prime} \cr s^{\prime}_{12}
c^{\prime}_{23} + c^{\prime}_{12}
s^{\prime}_{13} s^{\prime}_{23} e^{{\rm i} \delta^\prime}
& -c^{\prime}_{12} c^{\prime}_{23} +
s^{\prime}_{12} s^{\prime}_{13} s^{\prime}_{23} e^{{\rm i} \delta^\prime}
& -c^{\prime}_{13} s^{\prime}_{23} \cr s^{\prime}_{12} s^{\prime}_{23}
- c^{\prime}_{12} s^{\prime}_{13} c^{\prime}_{23}
e^{{\rm i} \delta^\prime} & -c^{\prime}_{12} s^{\prime}_{23}
- s^{\prime}_{12} s^{\prime}_{13}
c^{\prime}_{23} e^{{\rm i} \delta^\prime} & c^{\prime}_{13}
c^{\prime}_{23} \cr \end{matrix}
\right) \hspace{-0.17cm}
\left( \begin{matrix}
1 & 0 & 0 \cr 0 & e^{{\rm i} \phi} & 0 \cr 0 & 0 & e^{{\rm i} \varphi}
\cr \end{matrix} \right)
\end{eqnarray}
via the transformation $U^{\prime\dagger} M^{}_\nu U^{\prime *}
= \widehat{M}^{\prime}_\nu$, where
\begin{eqnarray}
\theta^{\prime}_{23} = \frac{\pi}{4} \; , ~~~ \delta^\prime
= \frac{\pi}{2} ~ {\rm or} ~
\frac{3\pi}{2} \; , ~~~ \phi = 0 ~{\rm or} ~ \frac{\pi}{2} \; , ~~~
\varphi = 0 ~ {\rm or} ~ \frac{\pi}{2} \; .
\end{eqnarray}
Because of $U^\prime = U S^{-1}$, the locations of $\phi$ and $\varphi$ in
$U^\prime$ are different from those in $U$. We
conclude that the $\mu$-$\tau$ reflection symmetry of $M^{}_\nu$
leads to both $|U^{}_{\mu i}| = |U^{}_{\tau i}|$ and
$|U^{\prime}_{\mu i}| = |U^{\prime}_{\tau i}|$ (for $i=1,2,3$),
although $U$ and $U^\prime$ correspond to the different neutrino
mass orderings.

The generic relationships between the flavor mixing parameters
$(\theta^{}_{12}, \theta^{}_{13}, \theta^{}_{23}, \delta)$
in $U$ and their counterparts in $U^\prime$ are given by
\begin{eqnarray}
t^\prime_{12} &= & \left| \frac{U^\prime_{e2}}{U^\prime_{e1}}
\right| = \left| \frac{U^{}_{e1}}{U^{}_{e3}} \right| =
\frac{c^{}_{12} c^{}_{13}}{s^{}_{13}} \; ,
\nonumber \\
s^\prime_{13} & = & \left| U^\prime_{e3} \right| = \left| U^{}_{e2}
\right| = s^{}_{12} c^{}_{13} \; ,
\nonumber \\
t^\prime_{23} &= & \left| \frac{U^\prime_{\mu 3}}{U^\prime_{\tau 3}}
\right| = \left| \frac{U^{}_{\mu 2}}{U^{}_{\tau 2}} \right| =
\frac{\sqrt{c^2_{12} c^2_{23} - 2 c^{}_{12} s^{}_{12} s^{}_{13}
c^{}_{23} s^{}_{23} c^{}_{\delta} + s^2_{12} s^2_{13} s^2_{23}}}
{\sqrt{c^2_{12} s^2_{23} + 2 c^{}_{12} s^{}_{12} s^{}_{13} c^{}_{23}
s^{}_{23} c^{}_{\delta} + s^2_{12} s^2_{13} c^2_{23}}} \; ,
\nonumber \\
s^\prime_{\delta} & = & \frac{{\rm Im}\left( U^\prime_{e2}
U^\prime_{\mu 3} U^{\prime *}_{e 3} U^{\prime *}_{\mu 2}\right)}
{c^\prime_{12} s^\prime_{12} c^{\prime 2}_{13} s^\prime_{13}
c^\prime_{23} s^\prime_{23}} = \frac{{\rm Im}\left( U^{}_{e1}
U^{}_{\mu 2} U^*_{e 2} U^*_{\mu 1}\right)} {c^\prime_{12}
s^\prime_{12} c^{\prime 2}_{13} s^\prime_{13} c^\prime_{23}
s^\prime_{23}} = \frac{ c^{}_{12} s^{}_{12} c^2_{13} s^{}_{13}
c^{}_{23} s^{}_{23}} {c^\prime_{12} s^\prime_{12} c^{\prime 2}_{13}
s^\prime_{13} c^\prime_{23} s^\prime_{23}} s^{}_{\delta} \; ,
\end{eqnarray}
where $t^\prime_{ij} \equiv \tan\theta^\prime_{ij}$, $c^{}_{\delta}
\equiv \cos\delta$, $s^{}_{\delta} \equiv \sin\delta$ and
$s^\prime_{\delta} \equiv \sin\delta^\prime$. It is easy to use
Eq. (14) to check that $\theta^\prime_{23} = \theta^{}_{23} = \pi/4$
and $\delta^\prime = \delta = \pi/2$ or $3\pi/2$ hold in the
$\mu$-$\tau$ reflection symmetry limit.
\begin{table}[t]
\caption{The best-fit values and $3\sigma$ ranges of six neutrino
oscillation parameters obtained from a global fit of currently available
experimental data \cite{Lisi}.}
\vspace{0.3cm}
\centering
\renewcommand\arraystretch{1.2}
\begin{tabular}{ccccccccc} \hline\hline
&& \multicolumn{3}{c}{Normal neutrino mass ordering} &&
\multicolumn{3}{c}{Inverted neutrino mass ordering} \\ \hline
&& best-fit & & $3\sigma$ range & & best-fit & & $3\sigma$ range
\\ \hline
$\theta^{}_{12}$ && $33.02^\circ$ && $30^\circ$---$36.51^\circ$
&& $33.02^\circ$ && $30^\circ$---$36.51^\circ$ \\
$\theta^{}_{13}$ && $8.43^\circ$ && $7.92^\circ$---$8.91^\circ$
&& $8.45^\circ$ && $7.92^\circ$---$8.95^\circ$ \\
$\theta^{}_{23}$ && $40.69^\circ$ && $38.12^\circ$---$51.65^\circ$
&& $50.13^\circ$ && $38.29^\circ$---$52.89^\circ$ \\
$\delta$ && $248.4^\circ$ && $0^\circ$---$30.6^\circ$ $\oplus$
$136.8^{\circ}$---$360 ^{\circ}$ && $235.8^\circ$
&&  $0^\circ$---$27^\circ$ $\oplus$
$124.2^{\circ}$---$360 ^{\circ}$ \\
\hline \\ \vspace{-1.35cm} \\
$\displaystyle \frac{\Delta m^{2}_{21}}{10^{-5} ~{\rm eV}^2}$ && $7.37$
&&$6.93$---$7.96$ && $7.37$ && $6.93$---$7.96$ \\ \vspace{-0.4cm} \\
$\displaystyle \frac{\Delta m^{2}_{31}}{10^{-3} ~{\rm eV}^2}$ &&
$2.56$ && $2.45$---$2.69$ && $-2.47$ && $-2.59$---$-2.35$ \\
\vspace{-0.6cm} \\
\hline\hline
\end{tabular}
\vspace{0.2cm}
\end{table}
\begin{figure*}[t]
\centering\includegraphics[width=16cm]{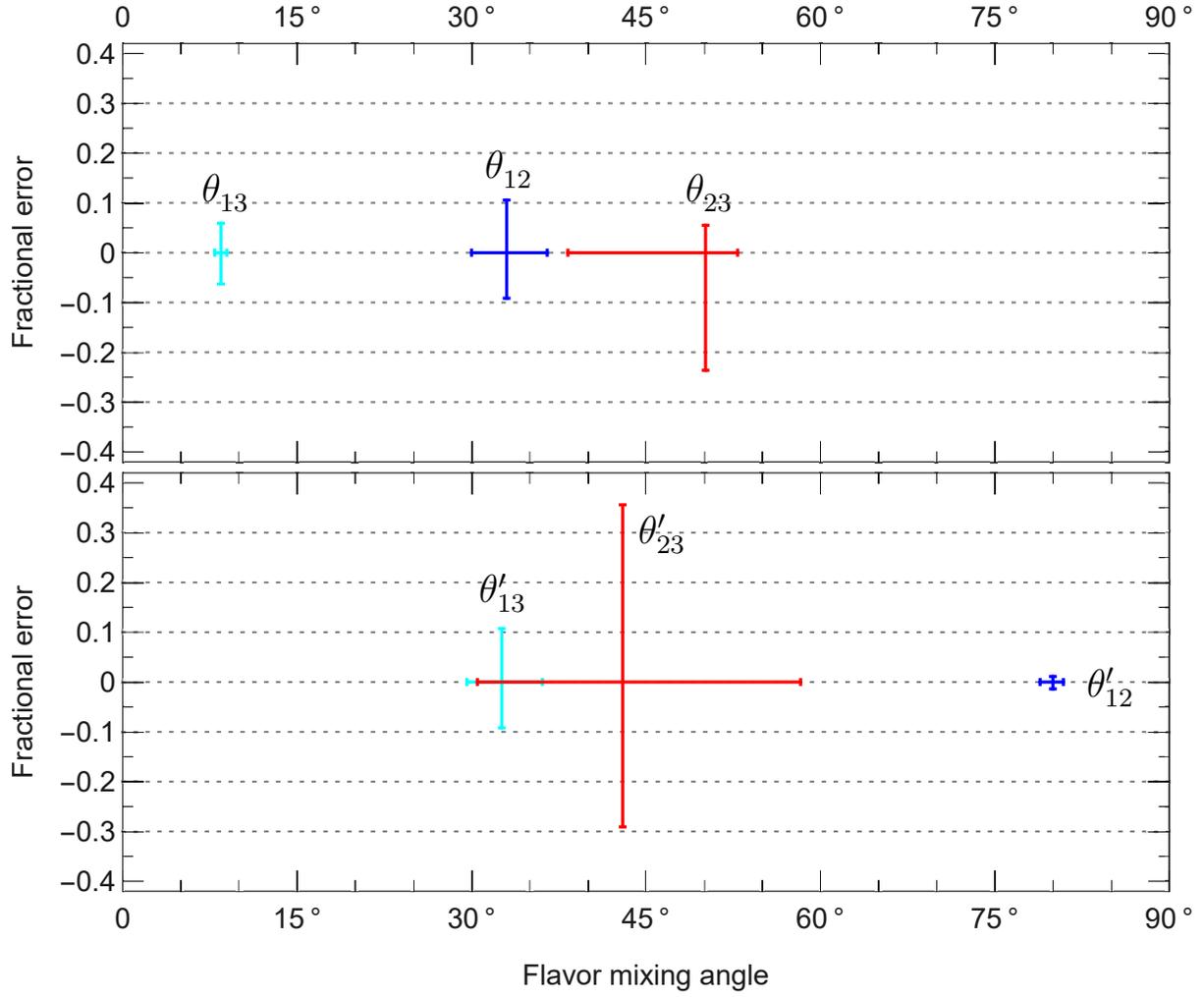}
\caption{An intuitive illustration of the numerical results of
$(\theta^{}_{12}, \theta^{}_{13}, \theta^{}_{23})$ and $(\theta^\prime_{12},
\theta^\prime_{13}, \theta^\prime_{23})$ obtained in Eq. (17)
for the IMO case.}
\end{figure*}

In Table 1 we list the latest global-fit results of six neutrino
oscillation parameters \cite{Lisi}.
It is obvious that the best-fit value of $\delta$ is close to
$\delta = 3\pi/2$ as indicated by the $\mu$-$\tau$ reflection symmetry
of $M^{}_\nu$. Hence we shall only take $\delta = 3\pi/2$ in the
remaining part of this section when discussing the
$\mu$-$\tau$ reflection symmetry limit. With the help of Table 1,
we compute the $3\sigma$ ranges of the elements of $U$ and
$U^\prime$ in the inverted mass ordering case. The results are
\begin{eqnarray}
|U| = \left(
\begin{matrix}
0.794 \to 0.858 & 0.494 \to 0.589 & 0.138 \to 0.156 \cr
0.194 \to 0.544 & 0.411 \to 0.728 & 0.612 \to 0.790 \cr
0.204 \to 0.550 & 0.425 \to 0.738 & 0.596 \to 0.777\cr
\end{matrix}
\right) \; ,
\end{eqnarray}
and
\begin{eqnarray}
|U^{\prime}| = \left(
\begin{matrix}
0.138 \to 0.156 & 0.794 \to 0.858 & 0.494 \to 0.589 \cr
0.612 \to 0.790 & 0.194 \to 0.544 & 0.411 \to 0.728 \cr
0.596 \to 0.777 & 0.204 \to 0.550 & 0.425 \to 0.738\cr
\end{matrix}
\right) \; .
\end{eqnarray}
To be more specific, the four flavor mixing parameters in the
standard parametrization of $U$ or $U^\prime$ read as
\begin{eqnarray}
\theta^{\prime}_{12} = 78.9^{\circ} \to 80.9^{\circ} \; ,   \quad
& \theta^{\prime}_{13} = 29.6^{\circ} \to 36.1^{\circ} \; , \quad
&\theta^{\prime}_{23} = 30.5^{\circ} \to 58.3^{\circ} \; ;
\nonumber\\
\theta^{}_{12} = 30.0^{\circ} \to 36.5^{\circ} \; ,   \quad
&\theta^{}_{13} = 7.9^{\circ} \to 8.9^{\circ} \; ,    \quad
&\theta^{}_{23} = 38.3^{\circ} \to 52.9^{\circ} \; ;
\end{eqnarray}
together with
\begin{eqnarray}
\delta \in [0^{\circ},27^{\circ}] \cup [124.2^{\circ},360^{\circ}]
\; ,  \qquad
\delta^{\prime} \in [0^{\circ},59.2^{\circ}] \cup [150.5^{\circ},360^{\circ}] \; .
\end{eqnarray}
We see that the value of $\theta^\prime_{12}$ is more than two times larger
than that of $\theta^{}_{12}$, and it is even not far
away from $90^\circ$. In addition, $\theta^\prime_{13}$ is about four times
larger than $\theta^{}_{13}$. But $\theta^\prime_{23}$ and $\theta^{}_{23}$
are roughly comparable in magnitude, so are $\delta^\prime$ and $\delta$.
Figure 3 gives a more intuitive comparison between the results of $\theta^{}_{ij}$ and
$\theta^\prime_{ij}$ obtained in Eq. (17), where
the fractional error bar of each flavor mixing angle is estimated from the difference
between the upper and lower bounds of its $3\sigma$ range divided by its
central value as listed in Table 1, and the corresponding errors translate from
$\theta^{}_{ij}$ to $\theta^\prime_{ij}$ via Eq. (14).

How about the parameters of flavor mixing and CP violation in a different
parametrization of $U$ and $U^\prime$?
To answer this question, we list all the nine angle-phase
parametrizations of $U$ and $U^\prime$ \cite{FX98} in Table 2 and calculate the
best-fit values of the corresponding four flavor mixing parameters
\footnote{Because the Majorana phases $\phi$ and $\varphi$ are
completely insensitive to neutrino oscillations and completely
unconstrained, here we only consider the Dirac CP-violating
phase $\delta$ or $\delta^\prime$ in our examples.}.
Some comments are in order.
\begin{itemize}
\item     The benchmark parametrization of $U$ or $U^\prime$ is
P3, which is actually consistent with the standard one given
in Eq. (10) or Eq. (12) up to a rearrangement of the sign
convention of $U^{}_{\mu i}$ (for $i=1,2,3$), a rearrangement of
the Dirac phase convention and a neglect
of the Majorana phase matrix. Starting from the best-fit values
of $\theta^{}_{12}$, $\theta^{}_{13}$, $\theta^{}_{23}$ and $\delta$
in Table 1, one may first determine the corresponding values
of $\theta^{\prime}_{12}$, $\theta^{\prime}_{13}$, $\theta^{\prime}_{23}$
and $\delta^\prime$ of $U^\prime$ in the form of P3, and then
translate the relevant results into the other eight parametrizations
of $U$ and $U^\prime$ as shown in Table 2.

\item     Among the nine parametrizations of $U$ and $U^\prime$,
only P1 \cite{FX97} allows all the three mixing angles of $U^\prime$
to be comparable in magnitude and smaller than $60^\circ$ for the given
inputs. In comparison, the outputs of four flavor mixing parameters
of $U^\prime$ in P2 (the Kobayashi-Maskawa parametrization \cite{CKM})
are quite similar to those in P3, with $\theta^\prime_{12} \sim 80^\circ$.
Another interesting parametrization of $U^\prime$, P5 \cite{Gerard},
contains $\theta^\prime_{12} \gtrsim 70^\circ$.
One can see that the parametrizations of $U^\prime$ under discussion may
involve one or two large mixing angles in most cases, as a straightforward
consequence of the reordering of the inverted neutrino mass hierarchy
$(\nu^{}_1, \nu^{}_2, \nu^{}_3) \to (\nu^\prime_1, \nu^\prime_2, \nu^\prime_3)
= (\nu^{}_3, \nu^{}_1, \nu^{}_2)$.
\end{itemize}
When the $\mu$-$\tau$ reflection symmetry is taken into account, as illustrated
in Table 3, we find that only P2, P3 and P7 are suitable in this connection
because they simply lead us to $\theta^{}_{23} = \theta^\prime_{23}
= \pi/4$ and $\delta = \delta^\prime = \pi/2$ or $3\pi/2$ as well
as $\phi = 0$ or $\pi/2$ and $\varphi = 0$ or $\pi/2$. Although
$|U^{}_{\mu i}| = |U^{}_{\tau i}|$ and
$|U^{\prime}_{\mu i}| = |U^{\prime}_{\tau i}|$ (for $i=1,2,3$)
also hold for the other six parametrizations of $U$ and $U^\prime$
in the $\mu$-$\tau$ reflection symmetry limit, the structures of
these parametrizations make themselves less interesting in describing
the phenomenology of neutrino oscillations.
\begin{table}[htb]
\centering \caption{The best-fit values of three flavor mixing
angles and the Dirac CP-violating phase for the inverted neutrino mass
ordering in each of the nine parametrizations of $U$ and $U^\prime$,
where $s^{}_{\theta^{}_{ij}} \equiv \sin
\theta^{}_{ij}$, $s^{}_{\vartheta^{}_{ij}} \equiv \sin
\vartheta^{}_{ij}$, $c^{}_{\theta^{}_{ij}} \equiv \cos
\theta^{}_{ij}$ and $c^{}_{\vartheta^{}_{ij}} \equiv \cos \vartheta^{}_{ij}$
(for $ij = 12, 13, 23$).}
\vspace{0.3cm}\footnotesize
\begin{tabular}[b]{lcc} \hline\hline
Parametrization  & $U$ & $U^{\prime}$\\ \hline \vspace{3mm}
\hspace{-0.2cm}$\begin{array}{l}
{\rm P1}: ~U=R_{12}^{}(\theta_{12}^{})
R_{23}^{}(\theta_{23}^{},\delta)
R_{12}^{-1}(\vartheta_{12}^{}) = \\ \begin{pmatrix}
s^{}_{\theta^{}_{12}} s^{}_{\vartheta_{12}^{}} c^{}_{\theta^{}_{23}} +
c^{}_{\theta_{12}^{}} c^{}_{\vartheta_{12}^{}} e^{-{\rm i} \delta}&
s^{}_{\theta^{}_{12}} c^{}_{\vartheta_{12}^{}} c^{}_{\theta^{}_{23}}
- c^{}_{\theta_{12}^{}} s^{}_{\vartheta_{12}^{}} e^{-{\rm i} \delta} &
s^{}_{\theta^{}_{12}} s^{}_{\theta^{}_{23}}\\c^{}_{\theta^{}_{12}}
s^{}_{\vartheta_{12}^{}} c^{}_{\theta^{}_{23}} - s^{}_{\theta_{12}^{}}
c^{}_{\vartheta_{12}^{}} e^{-{\rm i} \delta}&c^{}_{\theta^{}_{12}}
c^{}_{\vartheta_{12}^{}} c^{}_{\theta^{}_{23}} +
s^{}_{\theta_{12}^{}} s^{}_{\vartheta_{12}^{}} e^{-{\rm i} \delta} &
c^{}_{\theta^{}_{12}} s^{}_{\theta^{}_{23}}\\-s^{}_{\vartheta_{12}^{}}
s^{}_{\theta^{}_{23}}&-c^{}_{\vartheta_{12}^{}}
s^{}_{\theta^{}_{23}}&c^{}_{\theta^{}_{23} }\end{pmatrix}
\end{array}$&$\begin{array}{c} \theta^{}_{12}=11.0^{\circ} \\
\vartheta_{12}^{}=37.2^{\circ} \\ \theta^{}_{23}=50.6^{\circ} \\
\delta=308.3^{\circ} \end{array}$&$\begin{array}{c}
\theta^{\prime}_{12}=43.2^{\circ} \\ \vartheta^{\prime}_{12}=53.6^{\circ} \\
\theta^{\prime}_{23}=52.0^{\circ} \\ \delta^{\prime}=197.1^{\circ}
\end{array}$\\ \vspace{2.1mm}
\hspace{-0.2cm}$\begin{array}{l}
{\rm P2}: ~U=R_{23}^{}(\theta_{23}^{})
R_{12}^{}(\theta_{12}^{},\delta) R_{23}^{-1}(\vartheta_{23}^{}) = \\
\begin{pmatrix} c^{}_{\theta_{12}^{}}&s^{}_{\theta_{12}^{}}
c^{}_{\vartheta_{23}^{}}&-s^{}_{\theta^{}_{12}} s^{}_{
\vartheta_{23}^{}}\\ -s^{}_{\theta^{}_{12}} c^{}_{\theta^{}_{23}}
&c^{}_{\theta^{}_{12}} c^{}_{\theta_{23}^{}} c^{}_{\vartheta_{23}^{}} +
s^{}_{\theta_{23}^{}} s^{}_{\vartheta_{23}^{}} e^{-{\rm i}
\delta}&-c^{}_{\theta^{}_{12}} c^{}_{\theta_{23}^{}}
s^{}_{\vartheta_{23}^{}} + s^{}_{\theta_{23}^{}} c^{}_{\vartheta_{23}^{}}
e^{-{\rm i} \delta}  \\ s^{}_{\theta^{}_{12}}
s^{}_{\theta^{}_{23}}&-c^{}_{\theta^{}_{12}} s^{}_{\theta_{23}^{}}
c^{}_{\vartheta_{23}^{}} + c^{}_{\theta_{23}^{}} s^{}_{\vartheta_{23}^{}}
e^{-{\rm i} \delta}&c^{}_{\theta^{}_{12}} s^{}_{\theta_{23}^{}}
s^{}_{\vartheta_{23}^{}} + c^{}_{\theta_{23}^{}} c^{}_{\vartheta_{23}^{}}
e^{-{\rm i} \delta} \\\end{pmatrix} \end{array}$&$\begin{array}{c}
\theta^{}_{12}=34.0^{\circ} \\ \theta_{23}^{}=56.7^{\circ} \\
\vartheta_{23}^{}=15.3^{\circ} \\ \delta=297.4^{\circ}
\end{array}$&$\begin{array}{c} \theta^{\prime}_{12}=81.5^{\circ} \\
\theta^{\prime}_{23}=39.9^{\circ} \\ \vartheta^{\prime}_{23}=33.0^{\circ} \\
\delta^{\prime}=235.8^{\circ} \end{array}$\\  \vspace{2.1mm}
\hspace{-0.2cm}$\begin{array}{l}
{\rm P3}: ~U=R_{23}^{}(\theta_{23}^{})
R_{31}^{}(\theta_{13}^{},\delta) R_{12}^{}(\theta_{12}^{}) = \\
\begin{pmatrix} c^{}_{\theta_{12}^{}} c^{}_{\theta_{13}^{}} &
s^{}_{\theta_{12}^{}} c^{}_{\theta^{}_{13}} & s^{}_{\theta^{}_{13}}\\
-c^{}_{\theta^{}_{12}} s^{}_{\theta_{13}^{}} s^{}_{\theta^{}_{23}} -
s^{}_{\theta_{12}^{}} c^{}_{\theta_{23}^{}} e^{-{\rm i} \delta} &
-s^{}_{\theta^{}_{12}} s^{}_{\theta_{13}^{}} s^{}_{\theta^{}_{23}} +
c^{}_{\theta_{12}^{}} c^{}_{\theta_{23}^{}} e^{-{\rm i} \delta} &
c^{}_{\theta^{}_{13}} s^{}_{\theta^{}_{23}}  \\
-c^{}_{\theta^{}_{12}} s^{}_{\theta_{13}^{}} c^{}_{\theta^{}_{23}} +
s^{}_{\theta_{12}^{}} s^{}_{\theta_{23}^{}} e^{-{\rm i} \delta} &
-s^{}_{\theta^{}_{12}} s^{}_{\theta_{13}^{}} c^{}_{\theta^{}_{23}} -
c^{}_{\theta_{12}^{}} s^{}_{\theta_{23}^{}} e^{-{\rm i} \delta}
& c^{}_{\theta^{}_{13}} c^{}_{\theta^{}_{23}} \\\end{pmatrix}
\end{array}$&$\begin{array}{c}
\theta^{}_{12}=33.0^{\circ} \\ \theta_{13}^{}=8.5^{\circ} \\
\theta^{}_{23}=50.1^{\circ} \\ \delta=235.8^{\circ}
\end{array}$&$\begin{array}{c}
\theta^{\prime}_{12}=80.0^{\circ} \\ \theta^{\prime}_{13}=32.6^{\circ} \\
\theta^{\prime}_{23}=43.0^{\circ} \\ \delta^{\prime}=305.3^{\circ}
\end{array}$\\ \vspace{2.1mm}
\hspace{-0.2cm}$\begin{array}{l}
{\rm P4}: ~U=R_{12}^{}(\theta_{12}^{})
R_{31}^{}(\theta_{13}^{},\delta)
R_{23}^{-1}(\theta_{23}^{}) = \\ \begin{pmatrix}
c^{}_{\theta_{12}^{}} c^{}_{\theta_{13}^{}} & c^{}_{\theta^{}_{12}}
s^{}_{\theta_{13}^{}} s^{}_{\theta^{}_{23}} + s^{}_{\theta_{12}^{}}
c^{}_{\theta_{23}^{}} e^{-{\rm i} \delta}  & c^{}_{\theta^{}_{12}}
s^{}_{\theta_{13}^{}} c^{}_{\theta^{}_{23}} - s^{}_{\theta_{12}^{}}
s^{}_{\theta_{23}^{}} e^{-{\rm i} \delta} \\
-s^{}_{\theta^{}_{12}} c^{}_{\theta_{13}^{}} &  -s^{}_{\theta^{}_{12}}
s^{}_{\theta_{13}^{}} s^{}_{\theta^{}_{23}} + c^{}_{\theta_{12}^{}}
c^{}_{\theta_{23}^{}} e^{-{\rm i} \delta} &  -s^{}_{\theta^{}_{12}}
s^{}_{\theta_{13}^{}} c^{}_{\theta^{}_{23}} - c^{}_{\theta_{12}^{}}
s^{}_{\theta_{23}^{}} e^{-{\rm i} \delta}   \\ -s^{}_{\theta^{}_{13}} &
c^{}_{\theta_{13}^{}}s^{}_{\theta^{}_{23}}
& c^{}_{\theta^{}_{13}} c^{}_{\theta^{}_{23}} \\\end{pmatrix}
\end{array}$&$\begin{array}{c}
\theta^{}_{12}=20.3^{\circ} \\ \theta_{13}^{}=27.9^{\circ} \\
\theta^{}_{23}=44.2^{\circ} \\ \delta=333.2^{\circ}
\end{array}$&$\begin{array}{c}
\theta^{\prime}_{12}=79.0^{\circ} \\ \theta^{\prime}_{13}=39.4^{\circ} \\
\theta^{\prime}_{23}=37.2^{\circ} \\ \delta^{\prime}=308.3^{\circ}
\end{array}$\\	\vspace{2.1mm}
\hspace{-0.2cm}$\begin{array}{l}
{\rm P5}: ~U=R_{31}^{}(\theta_{13}^{})
R_{12}^{}(\theta_{12}^{},\delta)
R_{31}^{-1}(\vartheta_{13}^{}) = \\ \begin{pmatrix}
c^{}_{\theta^{}_{12}} c^{}_{\theta_{13}^{}} c^{}_{\vartheta_{13}^{}} +
s^{}_{\theta_{13}^{}} s^{}_{\vartheta_{13}^{}} e^{-{\rm i} \delta} &
s^{}_{\theta^{}_{12}} c^{}_{\theta_{13}^{}} & -c^{}_{\theta^{}_{12}}
c^{}_{\theta_{13}^{}} s^{}_{\vartheta_{13}^{}} + s^{}_{\theta_{13}^{}}
c^{}_{\vartheta_{13}^{}} e^{-{\rm i} \delta} \\
-s^{}_{\theta^{}_{12}} c^{}_{\vartheta_{13}^{}} &  c^{}_{\theta^{}_{12}} &
s^{}_{\theta^{}_{12}} s^{}_{\vartheta_{13}^{}}\\ -c^{}_{\theta^{}_{12}}
s^{}_{\theta_{13}^{}} c^{}_{\vartheta_{13}^{}} + c^{}_{\theta_{13}^{}}
s^{}_{\vartheta_{13}^{}} e^{-{\rm i} \delta} & -s^{}_{\theta^{}_{12}}
s^{}_{\theta_{13}^{}}
& c^{}_{\theta^{}_{12}} s^{}_{\theta_{13}^{}} s^{}_{\vartheta_{13}^{}} +
c^{}_{\theta_{13}^{}} c^{}_{\vartheta_{13}^{}} e^{-{\rm i} \delta}
\\\end{pmatrix} \end{array}$&$\begin{array}{c}
\theta^{}_{12}=54.9^{\circ} \\ \theta_{13}^{}=48.8^{\circ} \\
\vartheta_{13}^{}=68.0^{\circ} \\ \delta=336.2^{\circ}
\end{array}$&$\begin{array}{c}
\theta^{\prime}_{12}=72.2^{\circ} \\ \theta^{\prime}_{13}=29.4^{\circ} \\
\vartheta^{\prime}_{13}=37.1^{\circ} \\ \delta^{\prime}=207.9^{\circ}
\end{array}$\\	\vspace{2.1mm}
\hspace{-0.2cm}$\begin{array}{l}
{\rm P6}: ~U=R_{12}^{}(\theta_{12}^{})
R_{23}^{}(\theta_{23}^{},\delta)
R_{31}^{}(\theta_{13}^{}) = \\ \begin{pmatrix}
-s^{}_{\theta^{}_{12}} s^{}_{\theta_{13}^{}} s^{}_{\theta^{}_{23}} +
c^{}_{\theta_{12}^{}} c^{}_{\theta_{13}^{}} e^{-{\rm i} \delta}  &
s^{}_{\theta^{}_{12}} c^{}_{\theta_{23}^{}} & s^{}_{\theta^{}_{12}}
c^{}_{\theta_{13}^{}} s^{}_{\theta^{}_{23}} + c^{}_{\theta_{12}^{}}
s^{}_{\theta_{13}^{}} e^{-{\rm i} \delta} \\
-c^{}_{\theta^{}_{12}} s^{}_{\theta_{13}^{}} s^{}_{\theta^{}_{23}} -
s^{}_{\theta_{12}^{}} c^{}_{\theta_{13}^{}} e^{-{\rm i} \delta} &
c^{}_{\theta^{}_{12}} c^{}_{\theta_{23}^{}} &  c^{}_{\theta^{}_{12}}
c^{}_{\theta_{13}^{}} s^{}_{\theta^{}_{23}} - s^{}_{\theta_{12}^{}}
s^{}_{\theta_{13}^{}} e^{-{\rm i} \delta}   \\ -s^{}_{\theta^{}_{13}}
c^{}_{\theta_{23}^{}} &  -s^{}_{\theta^{}_{23}}
& c^{}_{\theta^{}_{13}} c^{}_{\theta^{}_{23}} \\\end{pmatrix}
\end{array}$&$\begin{array}{c}
\theta^{}_{12}=43.2^{\circ} \\ \theta_{13}^{}=36.4^{\circ} \\
\theta^{}_{23}=38.0^{\circ} \\ \delta=197.1^{\circ}
\end{array}$&$\begin{array}{c}
\theta^{\prime}_{12}=69.7^{\circ} \\ \theta^{\prime}_{13}=45.8^{\circ} \\
\theta^{\prime}_{23}=27.9^{\circ} \\ \delta^{\prime}=333.2^{\circ}
\end{array}$\\	\vspace{2.1mm}
\hspace{-0.2cm}$\begin{array}{l}
{\rm P7}: ~U=R_{23}^{}(\theta_{23}^{})
R_{12}^{}(\theta_{12}^{},\delta)
R_{31}^{-1}(\theta_{13}^{}) = \\ \begin{pmatrix}
c^{}_{\theta_{12}^{}} c^{}_{\theta_{13}^{}} & s^{}_{\theta^{}_{12}} &
-c^{}_{\theta^{}_{12}} s^{}_{\theta_{13}^{}} \\
-s^{}_{\theta^{}_{12}} c^{}_{\theta_{13}^{}} c^{}_{\theta^{}_{23}} +
s^{}_{\theta_{13}^{}} s^{}_{\theta_{23}^{}} e^{-{\rm i} \delta} &
c^{}_{\theta^{}_{12}} c^{}_{\theta_{23}^{}} &  s^{}_{\theta^{}_{12}}
s^{}_{\theta_{13}^{}} c^{}_{\theta^{}_{23}} + c^{}_{\theta_{13}^{}}
s^{}_{\theta_{23}^{}} e^{-{\rm i} \delta}   \\ s^{}_{\theta^{}_{12}}
c^{}_{\theta_{13}^{}} s^{}_{\theta^{}_{23}} + s^{}_{\theta_{13}^{}}
c^{}_{\theta_{23}^{}} e^{-{\rm i} \delta} &  -c^{}_{\theta^{}_{12}}
s^{}_{\theta_{23}^{}} &  -s^{}_{\theta^{}_{12}} s^{}_{\theta_{13}^{}}
s^{}_{\theta^{}_{23}} + c^{}_{\theta_{13}^{}} c^{}_{\theta_{23}^{}}
e^{-{\rm i} \delta}  \\\end{pmatrix} \end{array}$&$\begin{array}{c}
\theta^{}_{12}=32.6^{\circ} \\ \theta_{13}^{}=10.0^{\circ} \\
\theta^{}_{23}=47.0^{\circ} \\ \delta=305.3^{\circ}
\end{array}$&$\begin{array}{c}
\theta^{\prime}_{12}=56.0^{\circ} \\ \theta^{\prime}_{13}=74.7^{\circ} \\
\theta^{\prime}_{23}=56.7^{\circ} \\ \delta^{\prime}=297.4^{\circ}
\end{array}$\\	\vspace{2.1mm}
\hspace{-0.2cm}$\begin{array}{l}
{\rm P8}: ~U=R_{31}^{}(\theta_{13}^{})
R_{12}^{}(\theta_{12}^{},\delta)
R_{23}^{}(\theta_{23}^{}) = \\ \begin{pmatrix}
c^{}_{\theta_{12}^{}} c^{}_{\theta_{13}^{}} & s^{}_{\theta^{}_{12}}
c^{}_{\theta_{13}^{}} c^{}_{\theta^{}_{23}} - s^{}_{\theta_{13}^{}}
s^{}_{\theta_{23}^{}} e^{-{\rm i} \delta}  & s^{}_{\theta^{}_{12}}
c^{}_{\theta_{13}^{}} s^{}_{\theta^{}_{23}} + s^{}_{\theta_{13}^{}}
c^{}_{\theta_{23}^{}} e^{-{\rm i} \delta} \\
-s^{}_{\theta^{}_{12}} &  c^{}_{\theta^{}_{12}} c^{}_{\theta_{23}^{}}  &
c^{}_{\theta^{}_{12}} s^{}_{\theta_{23}^{}}  \\ -c^{}_{\theta^{}_{12}}
s^{}_{\theta^{}_{13}} &  -s^{}_{\theta^{}_{12}} s^{}_{\theta_{13}^{}}
c^{}_{\theta^{}_{23}} - c^{}_{\theta_{13}^{}} s^{}_{\theta_{23}^{}}
e^{-{\rm i} \delta} & -s^{}_{\theta^{}_{12}} s^{}_{\theta_{13}^{}}
s^{}_{\theta^{}_{23}} + c^{}_{\theta_{13}^{}} c^{}_{\theta_{23}^{}}
e^{-{\rm i} \delta} \\\end{pmatrix} \end{array}$&$\begin{array}{c}
\theta^{}_{12}=17.8^{\circ} \\ \theta_{13}^{}=29.4^{\circ} \\
\theta^{}_{23}=52.9^{\circ} \\ \delta=207.9^{\circ}
\end{array}$&$\begin{array}{c}
\theta^{\prime}_{12}=49.4^{\circ} \\ \theta^{\prime}_{13}=77.0^{\circ} \\
\theta^{\prime}_{23}=61.9^{\circ} \\ \delta^{\prime}=294.5^{\circ}
\end{array}$\\	\vspace{2.1mm}
\hspace{-0.2cm}$\begin{array}{l}
{\rm P9}: ~U=R_{31}^{}(\theta_{13}^{})
R_{23}^{}(\theta_{23}^{},\delta)
R_{12}^{-1}(\theta_{12}^{}) = \\ \begin{pmatrix}
-s^{}_{\theta^{}_{12}} s^{}_{\theta_{13}^{}} s^{}_{\theta^{}_{23}} + c^{}_{\theta_{12}^{}} c^{}_{\theta_{13}^{}} e^{-{\rm i} \delta}
& -c^{}_{\theta^{}_{12}} s^{}_{\theta_{13}^{}} s^{}_{\theta^{}_{23}} -
s^{}_{\theta_{12}^{}} c^{}_{\theta_{13}^{}} e^{-{\rm i} \delta}
& s^{}_{\theta^{}_{13}} c^{}_{\theta_{23}^{}} \\
s^{}_{\theta^{}_{12}} c^{}_{\theta_{23}^{}} &  c^{}_{\theta^{}_{12}}
c^{}_{\theta_{23}^{}} &  s^{}_{\theta^{}_{23}} \\-s^{}_{\theta^{}_{12}}
c^{}_{\theta_{13}^{}} s^{}_{\theta^{}_{23}} - c^{}_{\theta_{12}^{}}
s^{}_{\theta_{13}^{}} e^{-{\rm i} \delta}  & -c^{}_{\theta^{}_{12}}
c^{}_{\theta_{13}^{}} s^{}_{\theta^{}_{23}} + s^{}_{\theta_{12}^{}}
s^{}_{\theta_{13}^{}} e^{-{\rm i} \delta}&  c^{}_{\theta_{13}^{}}
c^{}_{\theta^{}_{23}} \\\end{pmatrix} \end{array}$&$\begin{array}{c}
\theta^{}_{12}=28.1^{\circ} \\ \theta_{13}^{}=13.0^{\circ} \\
\theta^{}_{23}=49.4^{\circ} \\ \delta=294.5^{\circ}
\end{array}$&$\begin{array}{c} \theta^{\prime}_{12}=68.0^{\circ} \\
\theta^{\prime}_{13}=41.2^{\circ} \\ \theta^{\prime}_{23}=35.1^{\circ} \\
\delta^{\prime}=336.2^{\circ}
\end{array}$\\ \hline\hline
\end{tabular}
\end{table}
\begin{table}[htb]
\centering \caption{The best-fit values of $\theta^{}_{12}$ and
$\theta^{}_{13}$ together with $\theta_{23}^{}=\pi/4$
and $\delta=3\pi/2$ for P3 in the $\mu$-$\tau$ reflection symmetry limit
\cite{Zhou} and in the inverted neutrino mass ordering, and their variations
for the other eight parametrizations in the same case.}
\vspace{0.05cm}\footnotesize
\begin{tabular}[t]{lcc} \hline\hline
Parametrization  & $U$ & $U^{\prime}$\\ \hline \vspace{3mm}
\hspace{-0.2cm}$\begin{array}{l}
{\rm P1}: ~U=R_{12}^{}(\theta_{12}^{})
R_{23}^{}(\theta_{23}^{},\delta)
R_{12}^{-1}(\vartheta_{12}^{}) = \\ \begin{pmatrix}
s^{}_{\theta^{}_{12}} s^{}_{\vartheta_{12}^{}} c^{}_{\theta^{}_{23}} +
c^{}_{\theta_{12}^{}} c^{}_{\vartheta_{12}^{}} e^{-{\rm i} \delta}&
s^{}_{\theta^{}_{12}} c^{}_{\vartheta_{12}^{}} c^{}_{\theta^{}_{23}} -
c^{}_{\theta_{12}^{}} s^{}_{\vartheta_{12}^{}} e^{-{\rm i} \delta} &
s^{}_{\theta^{}_{12}} s^{}_{\theta^{}_{23}}\\c^{}_{\theta^{}_{12}}
s^{}_{\vartheta_{12}^{}} c^{}_{\theta^{}_{23}} - s^{}_{\theta_{12}^{}}
c^{}_{\vartheta_{12}^{}} e^{-{\rm i} \delta}&c^{}_{\theta^{}_{12}}
c^{}_{\vartheta_{12}^{}} c^{}_{\theta^{}_{23}} + s^{}_{\theta_{12}^{}}
s^{}_{\vartheta_{12}^{}} e^{-{\rm i} \delta} & c^{}_{\theta^{}_{12}}
s^{}_{\theta^{}_{23}}\\-s^{}_{\vartheta_{12}^{}}
s^{}_{\theta^{}_{23}}&-c^{}_{\vartheta_{12}^{}}
s^{}_{\theta^{}_{23}}&c^{}_{\theta^{}_{23} }\end{pmatrix}
\end{array}$&$\begin{array}{c}
\theta^{}_{12}=11.9^{\circ} \\ \vartheta_{12}^{}=33.6^{\circ} \\
\theta^{}_{23}=45.6^{\circ} \\ \delta=277.3^{\circ} \end{array}$ &
$\begin{array}{c} \theta^{\prime}_{12}=42.1^{\circ} \\
\vartheta^{\prime}_{12}=60.5^{\circ} \\ \theta^{\prime}_{23} =
53.4^{\circ} \\ \delta^{\prime}=203.7^{\circ} \end{array}$\\
\vspace{2.1mm}
\hspace{-0.2cm}$\begin{array}{l}
{\rm P2}: ~U=R_{23}^{}(\theta_{23}^{})
R_{12}^{}(\theta_{12}^{},\delta)
R_{23}^{-1}(\vartheta_{23}^{}) = \\ \begin{pmatrix}
c^{}_{\theta_{12}^{}}&s^{}_{\theta_{12}^{}}
c^{}_{\vartheta_{23}^{}}&-s^{}_{\theta^{}_{12}} s^{}_{\vartheta_{23}^{}}\\
-s^{}_{\theta^{}_{12}} c^{}_{\theta^{}_{23}}&c^{}_{\theta^{}_{12}}
c^{}_{\theta_{23}^{}} c^{}_{\vartheta_{23}^{}} + s^{}_{\theta_{23}^{}}
s^{}_{\vartheta_{23}^{}} e^{-{\rm i} \delta}&-c^{}_{\theta^{}_{12}}
c^{}_{\theta_{23}^{}} s^{}_{\vartheta_{23}^{}} + s^{}_{\theta_{23}^{}}
c^{}_{\vartheta_{23}^{}} e^{-{\rm i} \delta}  \\ s^{}_{\theta^{}_{12}}
s^{}_{\theta^{}_{23}}&-c^{}_{\theta^{}_{12}} s^{}_{\theta_{23}^{}}
c^{}_{\vartheta_{23}^{}} + c^{}_{\theta_{23}^{}} s^{}_{\vartheta_{23}^{}}
e^{-{\rm i} \delta}&c^{}_{\theta^{}_{12}} s^{}_{\theta_{23}^{}}
s^{}_{\vartheta_{23}^{}} + c^{}_{\theta_{23}^{}} c^{}_{\vartheta_{23}^{}}
e^{-{\rm i} \delta} \\\end{pmatrix} \end{array}$&$\begin{array}{c}
\theta^{}_{12}=34.0^{\circ} \\ \theta_{23}^{}=45.0^{\circ} \\
\vartheta_{23}^{}=15.3^{\circ} \\ \delta=270.0^{\circ}
\end{array}$&$\begin{array}{c} \theta^{\prime}_{12}=81.5^{\circ} \\
\theta^{\prime}_{23}=45.0^{\circ} \\ \vartheta^{\prime}_{23}=33.0^{\circ} \\
\delta^{\prime}=270.0^{\circ} \end{array}$\\  \vspace{2.1mm}
\hspace{-0.2cm}$\begin{array}{l}
{\rm P3}: ~U=R_{23}^{}(\theta_{23}^{})
R_{31}^{}(\theta_{13}^{},\delta)
R_{12}^{}(\theta_{12}^{}) = \\ \begin{pmatrix}
c^{}_{\theta_{12}^{}} c^{}_{\theta_{13}^{}} & s^{}_{\theta_{12}^{}}
c^{}_{\theta^{}_{13}} & s^{}_{\theta^{}_{13}}\\
-c^{}_{\theta^{}_{12}} s^{}_{\theta_{13}^{}} s^{}_{\theta^{}_{23}} -
s^{}_{\theta_{12}^{}} c^{}_{\theta_{23}^{}} e^{-{\rm i} \delta} &
-s^{}_{\theta^{}_{12}} s^{}_{\theta_{13}^{}} s^{}_{\theta^{}_{23}} +
c^{}_{\theta_{12}^{}} c^{}_{\theta_{23}^{}} e^{-{\rm i} \delta} &
c^{}_{\theta^{}_{13}} s^{}_{\theta^{}_{23}}  \\ -c^{}_{\theta^{}_{12}}
s^{}_{\theta_{13}^{}} c^{}_{\theta^{}_{23}} + s^{}_{\theta_{12}^{}}
s^{}_{\theta_{23}^{}} e^{-{\rm i} \delta} & -s^{}_{\theta^{}_{12}}
s^{}_{\theta_{13}^{}} c^{}_{\theta^{}_{23}} - c^{}_{\theta_{12}^{}}
s^{}_{\theta_{23}^{}} e^{-{\rm i} \delta}
& c^{}_{\theta^{}_{13}} c^{}_{\theta^{}_{23}} \\\end{pmatrix}
\end{array}$&$\begin{array}{c}
\theta^{}_{12}=33.0^{\circ} \\ \theta_{13}^{}=8.5^{\circ} \\
\theta^{}_{23}=45.0^{\circ} \\ \delta=270.0^{\circ}
\end{array}$&$\begin{array}{c}
\theta^{\prime}_{12}=80.0^{\circ} \\ \theta^{\prime}_{13}=32.6^{\circ} \\
\theta^{\prime}_{23}=45.0^{\circ} \\ \delta^{\prime}=270.0^{\circ}
\end{array}$\\ \vspace{2.1mm}
\hspace{-0.2cm}$\begin{array}{l}
{\rm P4}: ~U=R_{12}^{}(\theta_{12}^{})
R_{31}^{}(\theta_{13}^{},\delta)
R_{23}^{-1}(\theta_{23}^{}) = \\ \begin{pmatrix}
c^{}_{\theta_{12}^{}} c^{}_{\theta_{13}^{}} & c^{}_{\theta^{}_{12}}
s^{}_{\theta_{13}^{}} s^{}_{\theta^{}_{23}} + s^{}_{\theta_{12}^{}}
c^{}_{\theta_{23}^{}} e^{-{\rm i} \delta}  & c^{}_{\theta^{}_{12}}
s^{}_{\theta_{13}^{}} c^{}_{\theta^{}_{23}} - s^{}_{\theta_{12}^{}}
s^{}_{\theta_{23}^{}} e^{-{\rm i} \delta} \\
-s^{}_{\theta^{}_{12}} c^{}_{\theta_{13}^{}} &  -s^{}_{\theta^{}_{12}}
s^{}_{\theta_{13}^{}} s^{}_{\theta^{}_{23}} + c^{}_{\theta_{12}^{}}
c^{}_{\theta_{23}^{}} e^{-{\rm i} \delta} &  -s^{}_{\theta^{}_{12}}
s^{}_{\theta_{13}^{}} c^{}_{\theta^{}_{23}} - c^{}_{\theta_{12}^{}}
s^{}_{\theta_{23}^{}} e^{-{\rm i} \delta}   \\ -s^{}_{\theta^{}_{13}} &
c^{}_{\theta_{13}^{}}s^{}_{\theta^{}_{23}}
& c^{}_{\theta^{}_{13}} c^{}_{\theta^{}_{23}} \\\end{pmatrix}
\end{array}$&$\begin{array}{c}
\theta^{}_{12}=25.5^{\circ} \\ \theta_{13}^{}=23.3^{\circ} \\
\theta^{}_{23}=40.4^{\circ} \\ \delta=329.1^{\circ}
\end{array}$&$\begin{array}{c}
\theta^{\prime}_{12}=78.1^{\circ} \\ \theta^{\prime}_{13}=44.4^{\circ} \\
\theta^{\prime}_{23}=33.6^{\circ} \\ \delta^{\prime}=277.3^{\circ}
\end{array}$\\	\vspace{2.1mm}
\hspace{-0.2cm}$\begin{array}{l}
{\rm P5}: ~U=R_{31}^{}(\theta_{13}^{})
R_{12}^{}(\theta_{12}^{},\delta)
R_{31}^{-1}(\vartheta_{13}^{}) = \\ \begin{pmatrix}
c^{}_{\theta^{}_{12}} c^{}_{\theta_{13}^{}} c^{}_{\vartheta_{13}^{}} +
s^{}_{\theta_{13}^{}} s^{}_{\vartheta_{13}^{}} e^{-{\rm i} \delta} &
s^{}_{\theta^{}_{12}} c^{}_{\theta_{13}^{}} & -c^{}_{\theta^{}_{12}}
c^{}_{\theta_{13}^{}} s^{}_{\vartheta_{13}^{}} + s^{}_{\theta_{13}^{}}
c^{}_{\vartheta_{13}^{}} e^{-{\rm i} \delta} \\
-s^{}_{\theta^{}_{12}} c^{}_{\vartheta_{13}^{}} &  c^{}_{\theta^{}_{12}} &
s^{}_{\theta^{}_{12}} s^{}_{\vartheta_{13}^{}}\\ -c^{}_{\theta^{}_{12}}
s^{}_{\theta_{13}^{}} c^{}_{\vartheta_{13}^{}} + c^{}_{\theta_{13}^{}}
s^{}_{\vartheta_{13}^{}} e^{-{\rm i} \delta} & -s^{}_{\theta^{}_{12}}
s^{}_{\theta_{13}^{}} & c^{}_{\theta^{}_{12}} s^{}_{\theta_{13}^{}}
s^{}_{\vartheta_{13}^{}} + c^{}_{\theta_{13}^{}} c^{}_{\vartheta_
{13}^{}} e^{-{\rm i} \delta}\\\end{pmatrix} \end{array}$&$\begin{array}{c}
\theta^{}_{12}=53.4^{\circ} \\ \theta_{13}^{}=47.9^{\circ} \\
\vartheta_{13}^{}=60.5^{\circ} \\ \delta=336.3^{\circ}
\end{array}$&$\begin{array}{c} \theta^{\prime}_{12}=66.7^{\circ} \\
\theta^{\prime}_{13}=25.5^{\circ} \\ \vartheta^{\prime}_{13}=
40.4^{\circ} \\ \delta^{\prime}=210.9^{\circ}
\end{array}$\\	\vspace{2.1mm}
\hspace{-0.2cm}$\begin{array}{l}
{\rm P6}: ~U=R_{12}^{}(\theta_{12}^{})
R_{23}^{}(\theta_{23}^{},\delta)
R_{31}^{}(\theta_{13}^{}) = \\ \begin{pmatrix}
-s^{}_{\theta^{}_{12}} s^{}_{\theta_{13}^{}} s^{}_{\theta^{}_{23}} +
c^{}_{\theta_{12}^{}} c^{}_{\theta_{13}^{}} e^{-{\rm i} \delta}  &
s^{}_{\theta^{}_{12}} c^{}_{\theta_{23}^{}} & s^{}_{\theta^{}_{12}}
c^{}_{\theta_{13}^{}} s^{}_{\theta^{}_{23}} + c^{}_{\theta_{12}^{}}
s^{}_{\theta_{13}^{}} e^{-{\rm i} \delta} \\
-c^{}_{\theta^{}_{12}} s^{}_{\theta_{13}^{}} s^{}_{\theta^{}_{23}} -
s^{}_{\theta_{12}^{}} c^{}_{\theta_{13}^{}} e^{-{\rm i} \delta} &
c^{}_{\theta^{}_{12}} c^{}_{\theta_{23}^{}} &  c^{}_{\theta^{}_{12}}
c^{}_{\theta_{13}^{}} s^{}_{\theta^{}_{23}} - s^{}_{\theta_{12}^{}}
s^{}_{\theta_{13}^{}} e^{-{\rm i} \delta}   \\ -s^{}_{\theta^{}_{13}}
c^{}_{\theta_{23}^{}} &  -s^{}_{\theta^{}_{23}}
& c^{}_{\theta^{}_{13}} c^{}_{\theta^{}_{23}} \\\end{pmatrix}
\end{array}$&$\begin{array}{c}
\theta^{}_{12}=42.1^{\circ} \\ \theta_{13}^{}=29.5^{\circ} \\
\theta^{}_{23}=36.6^{\circ} \\ \delta=203.7^{\circ}
\end{array}$&$\begin{array}{c}
\theta^{\prime}_{12}=64.5^{\circ} \\ \theta^{\prime}_{13}=49.6^{\circ} \\
\theta^{\prime}_{23}=23.3^{\circ} \\ \delta^{\prime}=329.1^{\circ}
\end{array}$\\	\vspace{2.1mm}
\hspace{-0.2cm}$\begin{array}{l}
{\rm P7}: ~U=R_{23}^{}(\theta_{23}^{})
R_{12}^{}(\theta_{12}^{},\delta)
R_{31}^{-1}(\theta_{13}^{}) = \\ \begin{pmatrix}
c^{}_{\theta_{12}^{}} c^{}_{\theta_{13}^{}} & s^{}_{\theta^{}_{12}} &
-c^{}_{\theta^{}_{12}} s^{}_{\theta_{13}^{}} \\
-s^{}_{\theta^{}_{12}} c^{}_{\theta_{13}^{}} c^{}_{\theta^{}_{23}} +
s^{}_{\theta_{13}^{}} s^{}_{\theta_{23}^{}} e^{-{\rm i} \delta} &
c^{}_{\theta^{}_{12}} c^{}_{\theta_{23}^{}} &  s^{}_{\theta^{}_{12}}
s^{}_{\theta_{13}^{}} c^{}_{\theta^{}_{23}} + c^{}_{\theta_{13}^{}}
s^{}_{\theta_{23}^{}} e^{-{\rm i} \delta}   \\ s^{}_{\theta^{}_{12}}
c^{}_{\theta_{13}^{}} s^{}_{\theta^{}_{23}} + s^{}_{\theta_{13}^{}}
c^{}_{\theta_{23}^{}} e^{-{\rm i} \delta} &  -c^{}_{\theta^{}_{12}}
s^{}_{\theta_{23}^{}} &  -s^{}_{\theta^{}_{12}} s^{}_{\theta_{13}^{}}
s^{}_{\theta^{}_{23}} + c^{}_{\theta_{13}^{}} c^{}_{\theta_{23}^{}}
e^{-{\rm i} \delta}  \\\end{pmatrix} \end{array}$&$\begin{array}{c}
\theta^{}_{12}=32.6^{\circ} \\ \theta_{13}^{}=10.0^{\circ} \\
\theta^{}_{23}=45.0^{\circ} \\ \delta=270.0^{\circ} \end{array}$&
$\begin{array}{c} \theta^{\prime}_{12}=56.0^{\circ} \\ \theta^{\prime}
_{13}=74.7^{\circ} \\ \theta^{\prime}_{23}=45.0^{\circ} \\ \delta^{\prime}
=270.0^{\circ} \end{array}$\\	\vspace{2.1mm}
\hspace{-0.2cm}$\begin{array}{l}
{\rm P8}: ~U=R_{31}^{}(\theta_{13}^{})
R_{12}^{}(\theta_{12}^{},\delta)
R_{23}^{}(\theta_{23}^{}) = \\ \begin{pmatrix}
c^{}_{\theta_{12}^{}} c^{}_{\theta_{13}^{}} & s^{}_{\theta^{}_{12}}
c^{}_{\theta_{13}^{}} c^{}_{\theta^{}_{23}} - s^{}_{\theta_{13}^{}}
s^{}_{\theta_{23}^{}} e^{-{\rm i} \delta}  & s^{}_{\theta^{}_{12}}
c^{}_{\theta_{13}^{}} s^{}_{\theta^{}_{23}} + s^{}_{\theta_{13}^{}}
c^{}_{\theta_{23}^{}} e^{-{\rm i} \delta} \\
-s^{}_{\theta^{}_{12}} &  c^{}_{\theta^{}_{12}} c^{}_{\theta_{23}^{}}  &
c^{}_{\theta^{}_{12}} s^{}_{\theta_{23}^{}}  \\ -c^{}_{\theta^{}_{12}}
s^{}_{\theta^{}_{13}} &  -s^{}_{\theta^{}_{12}} s^{}_{\theta_{13}^{}}
c^{}_{\theta^{}_{23}} - c^{}_{\theta_{13}^{}} s^{}_{\theta_{23}^{}}
e^{-{\rm i} \delta} & -s^{}_{\theta^{}_{12}} s^{}_{\theta_{13}^{}}
s^{}_{\theta^{}_{23}} + c^{}_{\theta_{13}^{}} c^{}_{\theta_{23}^{}}
e^{-{\rm i} \delta} \\\end{pmatrix} \end{array}$&$\begin{array}{c}
\theta^{}_{12}=23.3^{\circ} \\ \theta_{13}^{}=25.5^{\circ} \\
\theta^{}_{23}=49.6^{\circ} \\ \delta=210.9^{\circ} \end{array}$&
$\begin{array}{c} \theta^{\prime}_{12}=44.4^{\circ} \\ \theta^{\prime}_{13}
=78.1^{\circ} \\ \theta^{\prime}_{23}=56.4^{\circ} \\ \delta^{\prime}
=262.7^{\circ} \end{array}$\\	\vspace{2.1mm}
\hspace{-0.2cm}$\begin{array}{l}
{\rm P9}: ~U=R_{31}^{}(\theta_{13}^{})
R_{23}^{}(\theta_{23}^{},\delta)
R_{12}^{-1}(\theta_{12}^{}) = \\ \begin{pmatrix}
-s^{}_{\theta^{}_{12}} s^{}_{\theta_{13}^{}} s^{}_{\theta^{}_{23}}
+ c^{}_{\theta_{12}^{}} c^{}_{\theta_{13}^{}} e^{-{\rm i} \delta}  &
-c^{}_{\theta^{}_{12}} s^{}_{\theta_{13}^{}} s^{}_{\theta^{}_{23}} -
s^{}_{\theta_{12}^{}} c^{}_{\theta_{13}^{}} e^{-{\rm i} \delta}
& s^{}_{\theta^{}_{13}} c^{}_{\theta_{23}^{}} \\
s^{}_{\theta^{}_{12}} c^{}_{\theta_{23}^{}} &  c^{}_{\theta^{}_{12}}
c^{}_{\theta_{23}^{}} &  s^{}_{\theta^{}_{23}} \\-s^{}_{\theta^{}_{12}}
c^{}_{\theta_{13}^{}} s^{}_{\theta^{}_{23}} - c^{}_{\theta_{12}^{}}
s^{}_{\theta_{13}^{}} e^{-{\rm i} \delta}  & -c^{}_{\theta^{}_{12}}
c^{}_{\theta_{13}^{}} s^{}_{\theta^{}_{23}} + s^{}_{\theta_{12}^{}}
s^{}_{\theta_{13}^{}} e^{-{\rm i} \delta}&  c^{}_{\theta_{13}^{}}
c^{}_{\theta^{}_{23}} \\\end{pmatrix} \end{array}$&$\begin{array}{c}
\theta^{}_{12}=33.6^{\circ} \\ \theta_{13}^{}=11.9^{\circ} \\
\theta^{}_{23}=44.4^{\circ} \\ \delta=262.7^{\circ} \end{array}$&
$\begin{array}{c} \theta^{\prime}_{12}=60.5^{\circ} \\ \theta^{\prime}
_{13}=42.1^{\circ} \\ \theta^{\prime}_{23}=36.6^{\circ} \\
\delta^{\prime}=336.3^{\circ} \end{array}$\\ \hline\hline
\end{tabular}
\end{table}

At this point let us briefly comment on the so-called quark-lepton
complementarity relations $\theta^{}_{12} + \theta^{\rm q}_{12}
= \pi/4$ \cite{QLC1} and $\theta^{}_{23} \pm \theta^{\rm q}_{23}
= \pi/4$ \cite{QLC2} in the standard parametrization of the PMNS
matrix $U$ and the CKM matrix $V$. Given $\theta^{\rm q}_{12} =
13.023^\circ \pm 0.038^\circ$, $\theta^{\rm q}_{13} =
0.201^{+0.009^\circ}_{-0.008^\circ}$, $\theta^{\rm q}_{23} =
2.361^{+0.063^\circ}_{-0.028^\circ}$ and $\delta^{}_{\rm q}
= 69.21^{+2.55^\circ}_{-4.59^\circ}$ extracted from current
experimental data on quark flavor mixing and CP violation \cite{PDG},
such relations seem acceptable as a phenomenological
conjecture
\footnote{Note that $U$ and $V$ are associated respectively with
$W^-$ and $W^+$, and hence it is more appropriate to compare between
$U$ and $V^\dagger$ in some sense \cite{Xing2013}. But $V$ itself
is approximately symmetric, so the three mixing angles of $V^\dagger$
are equal to those of $V$ to a good degree of accuracy.}.
One may follow a similar way to conjecture the relations like
$\theta^{\prime}_{12} + \theta^{\rm q}_{12}
= \pi/2$ and $\theta^{\prime}_{23} \pm \theta^{\rm q}_{23}
= \pi/4$ \cite{Xing2013} when reordering the inverted neutrino
mass hierarchy and taking account of the standard parametrization
of $U^\prime$, but they seem unlikely to shed light on the
underlying dynamics of flavor mixing.

\section{The $\mu$-$\tau$ symmetry breaking}

At the level of the PMNS matrix $U$ or $U^\prime$, a deviation
of $\theta^{}_{23}$ or $\theta^\prime_{23}$ from $\pi/4$ is the
so-called {\it octant} problem which is one of the important concerns
in current neutrino oscillation experiments. On the other hand,
a departure of $\delta$ or $\delta^\prime$ from $3\pi/2$ can be
referred to as the {\it quadrant} problem, provided the present best-fit
value of $\delta$ is essentially true. Both issues actually measure
the effects of $\mu$-$\tau$ reflection symmetry breaking, and hence
it makes sense to examine such effects with the help of Table 1.

At the level of the Majorana neutrino mass matrix $M^{}_\nu$,
the $\mu$-$\tau$ reflection symmetry is equivalent
to the four relations given in Eq. (9). That is why some authors
\cite{Rodejohann} have used the imaginary parts of $\langle m\rangle^{}_{ee}$
and $\langle m\rangle^{}_{\mu \tau}$ and the differences
$\langle m\rangle^{}_{e \mu} - \langle m\rangle^{*}_{e \tau}$
and $\langle m\rangle^{}_{\mu \mu} - \langle m\rangle^{*}_{\tau \tau}$
to measure the symmetry breaking effects. But these four
quantities depend on the phase convention of the charged-lepton
fields, and hence they are not fully physical.

Here we illustrate
the effects of $\mu$-$\tau$ reflection symmetry breaking in a
somewhat different way. We first calculate the profiles of
$|\langle m\rangle_{\alpha\beta}^{}|$
(for $\alpha, ~\beta = e, ~\mu, ~\tau$) versus the lightest neutrino
mass (either $m_1^{}$ in the normal ordering or $m_3^{}$ in
the inverted ordering) by inputting the $3\sigma$ ranges of
the relevant neutrino oscillation parameters listed in Table 1.
For each of the six profiles,
we figure out the much narrower areas fixed by the $\mu$-$\tau$ reflection
symmetry conditions $\theta^{}_{23} = \pi/4$, $\delta = \pi/2$
or $3\pi/2$, $\phi = 0$ or $\pi/2$ and $\varphi = 0$ or $\pi/2$
in the standard parametrization of $U$. Note that
the case of $\delta = \pi/2$ is equivalent to that of
$\delta = 3\pi/2$, because
$|\langle m \rangle^{}_{\alpha\beta}|$ only contains $\cos\delta$.
Therefore, we are left with four different situations with
$(\phi, \varphi) = (0, 0)$, $(\pi/2, 0)$, $(0, \pi/2)$ and
$(\pi/2, \pi/2)$, as shown in Figure 4
for the normal neutrino mass ordering or in Figure 5 for the
inverted mass ordering. Some discussions are in order.
\begin{itemize}
\item     In either Figure 4 or Figure 5, the $\mu$-$\tau$ reflection
symmetry reflected by the relationships $|\langle m \rangle _{e \mu}| =
|\langle m \rangle _{e \tau}|$ and $|\langle m \rangle _{\mu \mu}| =
|\langle m \rangle _{\tau \tau}|$ can clearly be seen. Given
$m^{}_1 \lesssim 0.1$ eV for the normal mass ordering or $m^{}_3 \lesssim
0.1$ eV for the inverted mass ordering, a reasonable upper bound extracted
from current cosmological data \cite{Planck}, we find that either
$|\langle m\rangle^{}_{\mu \tau}|$ in Figure 4 or $|\langle m\rangle^{}_{ee}|$
in Figure 5 can be most stringently constrained. Hence the effect of
$\mu$-$\tau$ reflection symmetry breaking for either of them is
relatively modest even at the $3\sigma$ level. In comparison,
the symmetry breaking effects are completely unconstrained at the
$3\sigma$ level for $|\langle m\rangle^{}_{e \mu}|$,
$|\langle m\rangle^{}_{\mu \mu}|$, $|\langle m\rangle^{}_{e \tau}|$
and $|\langle m\rangle^{}_{\tau \tau}|$ in the inverted mass
ordering case, as shown in Figure 5. For $m^{}_1 \lesssim
2\times 10^{-2}$ eV in the normal mass ordering case, Figure 4 shows
that $|\langle m\rangle^{}_{\mu \mu}|$ and $|\langle m\rangle^{}_{\tau \tau}|$
are well constrained (around a few $\times 10^{-2}$ eV), and the
corresponding $\mu$-$\tau$ reflection symmetry breaking effects
are quite modest.

\item     The size of $|\langle m\rangle^{}_{ee}|$ determines
the rates of the neutrinoless double-beta ($0\nu2\beta$) decays for some
even-even nuclei, which are the only feasible way at present to probe
the Majorana nature of massive neutrinos \cite{Furry}. It is obvious
that the inverted mass ordering is more favorable for this purpose,
because even the lower limit of $|\langle m\rangle^{}_{ee}|$ is
above $0.01$ eV, which is experimentally accessible in the foreseeable
future \cite{Giunti}. As for the normal neutrino mass ordering, what
interests us most is the possibility of significant cancellation among
the three components of $\langle m\rangle^{}_{ee}$ even in the $\mu$-$\tau$
reflection symmetry case (the green and red regions in the profile of
$|\langle m\rangle^{}_{ee}|$, as shown in Figure 4). To understand this
observation, we follow Ref. \cite{XZ2017} to express $|\langle m\rangle^{}_{ee}|$
in the following way in terms of two Majorana phases $\rho$ and $\sigma$:
\begin{eqnarray}
|\langle m\rangle^{}_{ee}| = \left| m^{}_1 |U^{}_{e 1}|^2 e^{{\rm i} \rho}
+ m^{}_2 |U^{}_{e 2}|^2 + m^{}_3 |U^{}_{e 3}|^2 e^{{\rm i} \sigma}\right| \; ,
\end{eqnarray}
where $\rho = 2 \left(\phi - \varphi\right)$ and $\sigma = -2 \left(\delta
+ \varphi\right)$ in the phase notations defined in Eq. (10). As pointed
out in Ref. \cite{XZ2017}, a necessary condition for
$|\langle m\rangle^{}_{ee}| \to 0$ is $\rho = \pm\pi$, which are equivalent
to $(\phi, \varphi) = (\pi/2, 0)$ (the green region) and $(\phi, \varphi) = (0, \pi/2)$
(the red region) in Figure 4, respectively. That is to say, among the four
possible combinations of the two Majorana phases $\phi$ and $\varphi$ in
the $\mu$-$\tau$ reflection symmetry limit, two of them are likely to make
$|\langle m\rangle^{}_{ee}|$ fall into a well (i.e., make the
rate of a $0\nu 2\beta$ decay too small to be observable) if the
value of $m^{}_1$ happens to lie in a ``wrong" region
\footnote{As for the inverted neutrino mass ordering,
$|\langle m\rangle^{}_{\mu \tau}| \to 0$ is also possible in the
$\mu$-$\tau$ reflection symmetry limit with
$\phi =\pi/2$ and $\varphi =0$ (the green region in Figure 5).}.
\end{itemize}
Of course, it is very difficult to pin down the effects of $\mu$-$\tau$
reflection symmetry breaking at the neutrino mass matrix level
unless the octant issue of $\theta^{}_{23}$ is experimentally resolved, the
absolute neutrino mass scale is fixed or at least further constrained,
and the three CP-violating phases are determined to an acceptable degree of accuracy.

But even the preliminary results obtained in Figures 4 and 5 can tell
us something useful. For example, one may draw the conclusion that
the well-known Fritzsch texture \cite{Fritzsch},
\begin{eqnarray}
M^{\rm (F)}_\nu = \left(\begin{matrix} {\bf 0} &
\langle m\rangle^{}_{e\mu} & {\bf 0} \cr \langle m\rangle^{}_{e\mu}  & {\bf 0}
& \langle m\rangle^{}_{\mu\tau} \cr
{\bf 0} & \langle m\rangle^{}_{\mu\tau} & \langle m\rangle^{}_{\tau\tau}
\cr\end{matrix}\right) \; ,
\end{eqnarray}
must not be applicable to the inverted neutrino mass ordering
even if the latter is relabeled to be normal. The reason is simply
that $\langle m\rangle^{}_{ee}$ is not only nonzero but also sizeable in
this case, as shown in Figure 5. Another interesting observation is
that there is strong tension between $M^{\rm (F)}_\nu$ and current
experimental data at the $3\sigma$ level even in the normal
mass ordering case, as Figure 4 shows that
the matrix elements $\langle m\rangle^{}_{ee}$ and
$\langle m\rangle^{}_{\mu\mu}$ cannot simultaneously vanish
\footnote{This result is in conflict with the previous
studies based on very preliminary neutrino oscillation data \cite{Xing2002}.}.

Another instructive example is the following two-zero texture of $M^{}_\nu$,
which is usually referred to as ``Pattern C" and favored with the
inverted neutrino mass ordering \cite{201,202}:
\begin{eqnarray}
M^{}_\nu = \left(\begin{matrix} \langle m\rangle^{}_{ee} &
\langle m\rangle^{}_{e\mu} & \langle m\rangle^{}_{e\tau} \cr
\langle m\rangle^{}_{e\mu}  & {\bf 0} & \langle m\rangle^{}_{\mu\tau} \cr
\langle m\rangle^{}_{e\tau} & \langle m\rangle^{}_{\mu\tau} & {\bf 0}
\cr\end{matrix}\right) \; .
\end{eqnarray}
A comparison of Eq. (21) with Figure 5 tells us that such a special
texture of $M^{}_\nu$ remains compatible with current neutrino
oscillation data at the $3\sigma$ level, but it does not respect the
$\mu$-$\tau$ reflection symmetry which definitely forbids
vanishing or very small (generously say, $\lesssim 10^{-3}$ eV)
values of $\langle m\rangle^{}_{\mu\mu}$
and $\langle m\rangle^{}_{\tau\tau}$.

Therefore, we expect that our model-independent results shown in Figures 4 and 5
can help for the model-building exercises in connection to the neutrino mass
spectrum, flavor mixing and CP violation, no matter whether one follows
the flavor symmetry approach or the texture-zero approach, or a
mixture between them.
\begin{figure*}[t]
\centering\includegraphics[width=16cm]{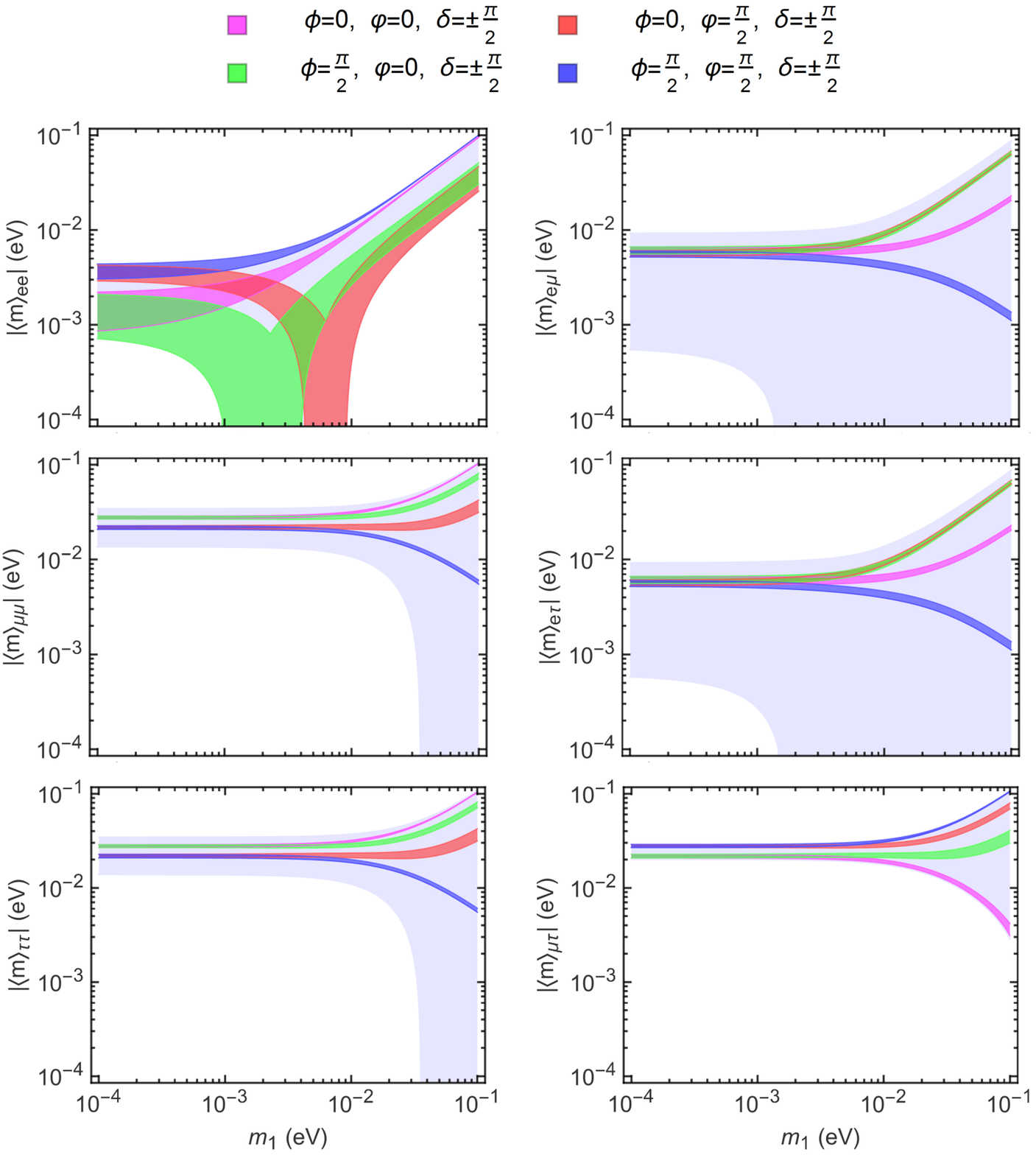}
\caption{The profiles of $|\langle m \rangle_{\alpha\beta}^{}|$ versus
the lightest neutrino mass $m_1^{}$ for the {\bf normal} neutrino mass
ordering, where the $3\sigma$ ranges of six neutrino oscillation
parameters \cite{Lisi} have been input. The pink, red, green and
blue regions are fixed by the $\mu$-$\tau$ reflection symmetry
with $\theta^{}_{23} = \pi/4$ and the special values of $\delta$,
$\phi$ and $\varphi$ listed on top of the figure.}
\end{figure*}
\begin{figure*}[t]
\centerline{\includegraphics[width=16cm]{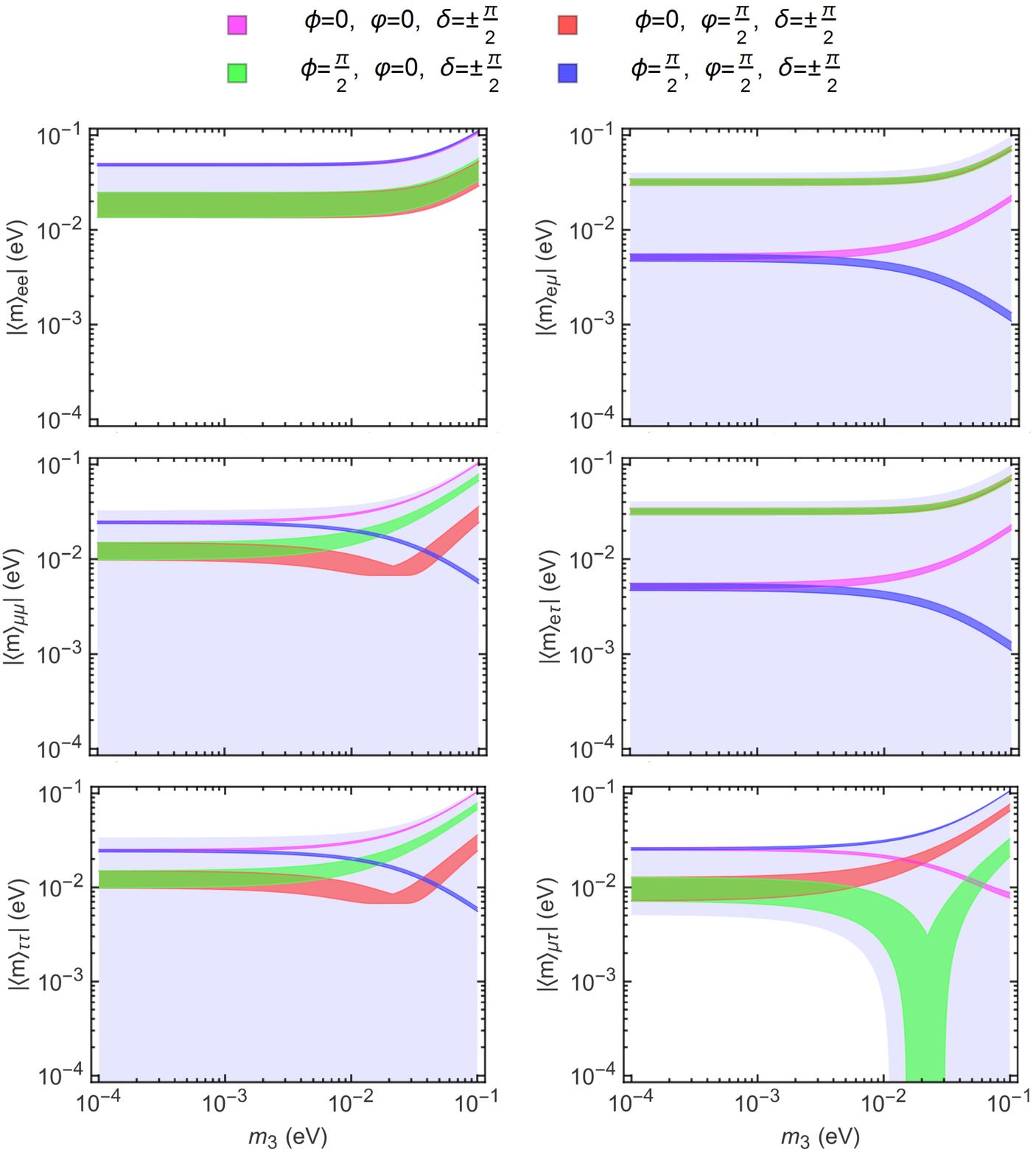}}
\caption{The profiles of $|\langle m \rangle_{\alpha\beta}^{}|$ versus
the lightest neutrino mass $m_3^{}$ for the {\bf inverted} neutrino mass
ordering, where the $3\sigma$ ranges of six neutrino oscillation
parameters \cite{Lisi} have been input. The pink, red, green and
blue regions are fixed by the $\mu$-$\tau$ reflection symmetry
with $\theta^{}_{23} = \pi/4$ and the special values of $\delta$,
$\phi$ and $\varphi$ listed on top of the figure.}
\end{figure*}

\section{Summary}

We have asked ourselves what to do if the neutrino mass spectrum
is finally found to be inverted. 
A straightforward and phenomenologically meaningful choice is
certainly to relabel $m^{}_3 < m^{}_1 < m^{}_2$ as
$m^\prime_1 < m^\prime_2 < m^\prime_3$, such that the latter is
as normal as the mass spectra of those charged leptons and quarks of
the same electrical charges. But
in this case the columns of the PMNS matrix $U$ must be reordered
accordingly, and the resulting pattern $U^\prime$ turns out to
involve an especially large mixing angle in the standard parametrization.
We have examined the other angle-phase parametrizations of
$U$ and $U^\prime$, and reached a similar observation. Given the fact that
the Majorana neutrino mass matrix $M^{}_\nu$ keeps unchanged in such
a mass relabeling exercise, we have considered the intriguing 
$\mu$-$\tau$ reflection symmetry to illustrate the
texture of $M^{}_\nu$ and the effects of its symmetry breaking
at the $3\sigma$ level by using current neutrino oscillation
data. Our model-independent results are expected to be helpful
for building specific neutrino mass models.

What is really behind the normal or inverted neutrino mass ordering
remains an open question, and
whether there is a definite correlation between the fermion mass
spectra and flavor mixing patterns is also a big puzzle.
The underlying flavor theory, which might be based on a certain
flavor symmetry and its spontaneous or explicit breaking mechanism,
may finally answer the above questions. To approach such a theory,
we are now following a phenomenological way from the bottom up.
Much more efforts in this connection are needed.

\vspace{0.5cm}

We would like to thank Y.F. Li, S. Luo, Z.H. Zhao and S. Zhou for
useful discussions. This work was supported in part by the National
Natural Science Foundation of China under grant No. 11135009 and No. 11375207.


\newpage
\begin{thebibliography}{99}

\bibitem{XZ} Z.Z. Xing and S. Zhou, {\it Neutrinos in
Particle Physics, Astronomy and Cosmology} (Zhejiang University
Press and Springer-Verlag, 2011).

\bibitem{PMNS} Z. Maki, M. Nakagawa, and S. Sakata,
Prog. Theor. Phys. {\bf 28}, 870 (1962); B. Pontecorvo, Sov. Phys.
JETP {\bf 26}, 984 (1968).

\bibitem{CKM} N. Cabibbo, Phys. Rev. Lett. {\bf 10}, 531 (1963);
M. Kobayashi and T. Maskawa, Prog. Theor. Phys. {\bf 49}, 652
(1973).

\bibitem{PDG} C. Patrignani {\it et al.} (Particle Data Group),
Chin. Phys. C {\bf 40}, 100001 (2016).

\bibitem{Hagiwara} See, e.g., M. Aoki {\it et al.},
Phys. Rev. D {\bf 67}, 093004 (2003).

\bibitem{Lisi} F. Capozzi, E. Di Valentino, E. Lisi, A. Marrone,
A. Melchiorri, and A. Palazzo, Phys. Rev. D {\bf 95}, 096014
(2017).

\bibitem{T2K} K. Abe {\it et al.} (T2K Collaboration), arXiv:1707.01048;
M. Hartz, {\it T2K Neutrino Oscillation Results with Data up to
2017 Summer}, talk given at the KEK Colloquium, August 4, 2017.

\bibitem{Schwetz} I. Esteban, M.C. Gonzalez-Garcia, M. Maltoni,
I. Martinez-Soler, and T. Schwetz, JHEP {\bf 1701}, 087 (2017);
A. Caldwell, A. Merle, O. Schulz, and M. Totzauer, arXiv:1705.01945.

\bibitem{XZZ} Z.Z. Xing, H. Zhang, and S. Zhou,
Phys. Rev. D {\bf 77}, 113016 (2008); Phys. Rev. D {\bf 86},
013013 (2012).

\bibitem{M} E. Majorana, Nuovo Cimento {\bf 14}, 171 (1937).

\bibitem{HS} P.F. Harrison and W.G. Scott, Phys. Lett. B {\bf 547},
219 (2002).

\bibitem{XZreview} For a recent review with extensive references, see:
Z.Z. Xing and Z.H. Zhao, Rept. Prog. Phys. {\bf 79},
076201 (2016).

\bibitem{FX98} H. Fritzsch and Z.Z. Xing, Phys. Rev. D {\bf 57},
594 (1998).

\bibitem{FX97} H. Fritzsch and Z.Z. Xing, Phys. Lett. B {\bf 413},
396 (1997).

\bibitem{Gerard} J.M. Gerard and Z.Z. Xing, Phys. Lett. B {\bf 713},
29 (2012); Z.Z. Xing, Chin. Phys. C {\bf 36}, 281 (2012).

\bibitem{Zhou} Z.Z. Xing and S. Zhou, Phys. Lett. B {\bf 666},
166 (2008).

\bibitem{QLC1} M. Raidal, Phys. Rev. Lett. {\bf 93}, 161801 (2004);
H. Minakata and A.Yu. Smirnov, Phys. Rev. D {\bf 70}, 073009 (2004).

\bibitem{QLC2} Z.Z. Xing, Phys. Lett. B {\bf 618}, 141 (2005).

\bibitem{Xing2013} Z.Z. Xing, arXiv:1309.2102 (unpublished).

\bibitem{Rodejohann} See, e.g., W. Grimus, A.S. Joshipura,
S. Kaneko, L. Lavoura, H. Sawanaka, and M. Tanimoto, Nucl.
Phys. B {\bf 713}, 151 (2005); Z.H. Zhao, arXiv:1703.04984;
W. Rodejohann and X.J. Xu, arXiv:1705.02027.

\bibitem{Planck} P.A.R. Ade {\it et al.} (Planck Collaboration),
Astron. Astrophys. {\bf 594}, A13 (2016).

\bibitem{Furry} W.H. Furry, Phys. Rev. D {\bf 56}, 1184 (1939).

\bibitem{Giunti} S.M. Bilenky and C. Giunti, Int. J. Mod. Phys.
A {\bf 30}, 1530001 (2015); and references therein.

\bibitem{XZ2017} Z.Z. Xing and Z.H. Zhao, Eur. Phys. J. C {\bf 77},
192 (2017). See also, Z.Z. Xing, Z.H. Zhao, and Y.L. Zhou,
Eur. Phys. J. C {\bf 75}, 423 (2015).

\bibitem{Fritzsch} H. Fritzsch, Phys. Lett. B {\bf 73}, 317 (1978);
Nucl. Phys. B {\bf 155}, 189 (1979).

\bibitem{Xing2002} Z.Z. Xing, Phys. Lett. B {\bf 550}, 178 (2002);
S. Zhou and Z.Z. Xing, Eur. Phys. J. C {\bf 38}, 495 (2005).

\bibitem{201} P.H. Frampton, S.L. Glashow, and D. Marfatia,
Phys. Lett. B {\bf 536}, 79 (2002).

\bibitem{202} Z.Z. Xing, Phys. Lett. B {\bf 530}, 159 (2002);
Phys. Lett. B {\bf 539}, 85 (2002); H. Fritzsch, Z.Z. Xing, and
S. Zhou, JHEP {\bf 1109}, 083 (2011); and references therein.

\end{thebibliography}
\end{document}